\documentclass[preprintnumbers,amsmath,amssymb,floatfix,11pt,prd,onecolumn,
superscriptaddress,nofootinbib]{revtex4}
\usepackage{latexsym}
\usepackage{epsfig}
\usepackage{epstopdf}
\usepackage{amssymb}
\usepackage{amsmath}
\usepackage{comment}
\usepackage{lineno,hyperref,pdflscape}
\usepackage{float}

\newcommand{\be}{\begin{equation}}
\newcommand{\ee}{\end{equation}}
\newcommand{\ben}{\begin{eqnarray}}
\newcommand{\een}{\end{eqnarray}}

\newcommand{\p}{\partial}

\begin{document}

\markboth{U. Camci, A. Yildirim}
{Noether gauge symmetry classes for pp-wave spacetimes}

\title{Noether gauge symmetry classes for pp-wave spacetimes}

\author{U. Camci}
\email{ucamci@akdeniz.edu.tr}\affiliation{Department of Physics, Akdeniz University, 07058 Antalya, Turkey}

\author{A. Yildirim}
\email{aydinyildirim@akdeniz.edu.tr}\affiliation{Department of Physics, Akdeniz University, 07058 Antalya, Turkey}

%%%%%%%%%%%%%%%%%%%%%%%%%%%%%%%%%%%%%%%%%%%%%%%%%%%%%%%%%%%%%%%%%%%%%%%%%%%%%%%
\begin{abstract}
The Noether gauge symmetries of geodesic Lagrangian for the pp-wave spacetimes are determined in each of the Noether gauge symmetry classes of the pp-wave spacetimes. It is shown that a type N pp-wave spacetime can admit at most \emph{three} Noether gauge symmetry, and furthermore the number of Noether gauge symmetries turn out to be \emph{four, five, six, seven} and \emph{eight}. We found that all conformally flat plane wave spacetimes admit the maximal, i.e. \emph{ten}, Noether gauge symmetry. Also it is found that if the pp-wave spacetime is non-conformally flat plane wave, then the number of Noether gauge symmetry is \emph{nine} or \emph{ten}. By means of the obtained Noether constants the search of the exact solutions of the geodesic equations for the pp-wave spacetimes is considered and we found new exact solutions of the geodesic equations in some special Noether gauge symmetry classes.

\end{abstract}

\maketitle

\textbf{Keywords:} pp-wave spacetime; plane wave spacetime; Noether gauge symmetry; geodesics.

\section{Introduction}
\label{INT}

In classical general relativity, the class of pp-wave spacetimes are the best-known and mathematically simplest class of solutions to Einstein's field equations \cite{kramer}. Ehlers and Kundt \cite{ehlers-kundt} investigated the properties of the vacuum ``plane-fronted gravitational waves with parallel rays", and they abbreviated this quoted term as ``pp-waves". They also determined the symmetry classes of pp-waves satisfying the vacuum field equations of general relativity. In vacuum, the pp-wave solution means that there is a geodesic null vector field. For a type N vacuum spacetime, this implies the existence of a covariantly constant null vector field, which is parallel to the vector field mentioned above. Podolsk$\acute{y}$ and Vesel$\acute{y}$ \cite{podolsky} investigated geodesics in non-homogeneous vacuum pp-wave spacetimes and demonstrated their chaotic behaviour by rigorous analytic and numerical methods.

We recall some facts on the conformal Killing symmetry. Let $M$ be a four-dimensional spacetime manifold with metric tensor $g$ of Lorentz signature. Any vector field ${\bf X}$ satisfying $\pounds_{\bf X} g = 2 \psi (x^a) g$ is said to be a conformal Killing vector (CKV) of the metric tensor $g$, where  $\psi (x^a)$ is a conformal factor. If $\psi_{;ab} \neq 0$, then
the CKV field is said to be {\it proper}. For other cases related with $\psi$, the vector field ${\bf X}$ is called the special conformal Killing vector (SCKV) field if $\psi_{;ab} = 0$, the homothetic Killing vector (HKV) field if $\psi_{, \, a} = 0$, e.g. $\psi$ is a constant on $M$,  and the Killing vector (KV) field if $\psi =0$. The set of all CKV (respectively SCKV, HKV and KV) form a finite-dimensional Lie algebra denoted by $\mathcal{C}$ (respectively $\mathcal{S}, \mathcal{H}$ and $\mathcal{G}$). The maximum dimension of the CKV algebra on $M$ is fifteen if $M$ is conformally flat, and it is seven if the spacetime is not conformally flat. The detailed study of the conformal symmetry properties of the null fluid pp-wave spacetimes is given in Refs.\cite{tupper2003} and \cite{kt}. These studies refine the isometry classification scheme given by Sippel and Goenner \cite{sg} and determine conformal symmetry classes for the pp-wave spacetimes. The general form for the CKV of a pp-wave spacetime and, in particular, the expressions for the HKV and SCKV in such spacetimes are determined by Maartens and Maharaj \cite{mm}. Hall \emph{et al} \cite{hall-hp} determined the CKV for general conformally flat pp-wave spacetime.

Let us remark on the Noether symmetry. Noether symmetries are interesting symmetries associated with differential equations possessing a Lagrangian, and they describe physical features of differential equations in terms of conservation laws (first integrals) admitted by them.
There exist two types Noether symmetry in the literature: the first one is the so-called Noether symmetry approach without gauge term \cite{capo00}-\cite{camci3} in which the Lie derivative of a given Lagrangian $L$ dragging along a vector field ${\bf Y}$ vanishes, i.e. $\pounds_{\bf Y} L = 0.$ The  second one is the so-called Noether gauge symmetry (NGS) approach \cite{camci2,ibrar,jamil} which is a generalization of the former Noether symmetry approach in the sense that Noether symmetry equation includes a gauge term, and it will be discussed in following section. It is noted here that the set of all NGS form a finite dimensional Lie algebra denoted by $\mathcal{N}$.

Lately Noether symmetries of the Lagrangian corresponding to the geodesic equations for some spacetimes have been calculated, and classified according to their symmetry generators \cite{feroze1}-\cite{tsamparlis2}. A connection between the KVs and NGSs of maximally symmetric spaces is examined by Feroze et al. \cite{feroze1}. For the spaces of different curvatures such as Bertotti-Robinson like spacetime, Feroze \cite{feroze2} has discussed and conjectured the existence of new conserved quantities. Later, Feroze and Hussain \cite{feroze3} have presented the form of new conserved quantities along with the proof of their conjecture. The NGSs of FRW spacetimes have studied by Tsamparlis and Paliathanasis \cite{tsamparlis1}. They have also examined the NGSs of class A Bianchi type homogeneous spacetimes with scalar field minimally coupled to gravity \cite{tsamparlis2}. Recently, Ali and Feroze \cite{feroze4} have provided a classification of plane symmetric static spacetimes considering Noether gauge symmetry approach. Using the approximate Noether symmetries the energy contents of some classes of colliding plane waves are calculated by Sharif and Waheed \cite{sharif2012}. Camci \cite{camci2014} has obtained the NGSs of geodesic Lagrangian $L$ and explicitly integrated the geodesic equations of motion for the corresponding stationary G\"{o}del-type spacetimes. In the earlier study, we investigated the Lie point symmetries of geodesic equations and Noether gauge symmetries of the geodesic Lagrangian for some classes of pp-wave spacetimes \cite{ua}. In this work, we aim to find the NGSs corresponding to geodesic Lagrangian for the general classes of pp-wave spacetimes.

In this study, we consider the line element of pp-wave spacetimes which can be written \cite{kramer} as
\begin{equation}
ds^2 = -2 H(u,y,z) du^2 -2 du dv + dy^2 + dz^2. \label{pp-wave}
\end{equation}
The pp-wave spacetime is vacuum if $H_{,yy} + H_{,zz} =0$ and {\it conformally flat} if $H_{,yy} = H_{,zz}$ and $H_{,yz}=0$, where a subscript together with comma implies partial derivative. Further, if the metric function can be put in the form
\begin{equation}
2 H(u,y,z) = A(u) y^2 + 2 B(u) y \, z + C(u) z^2, \label{plane-wave}
\end{equation}
then the spacetime is referred to as a plane wave spacetime. The form of the Riemann, Weyl and Ricci tensors for the pp-wave spacetime
\begin{eqnarray}
& & R_{uyuy} = H_{,yy}\, , \quad R_{uyuz} = H_{,yz}\, , \quad R_{uzuz} = H_{,zz} \label{Riemann} \\
& & C_{uyuy} = \frac{1}{2} (H_{,yy} - H_{,zz}) =-C_{uzuz}, \quad C_{uyuz} = H_{,yz}  \label{weyl} \\ & & R_{ab} = F k_a k_b \label{ricci}
\end{eqnarray}
where $F = H_{,yy}+ H_{,zz}$ and ${\bf k}$ is the null covariantly constant Killing vector which has the form $k^a = \delta^a_v, \,\, k_a = - \delta^u_a$. Therefore the Ricci scalar for pp-wave spacetimes vanishes. We point out here a fact that the Petrov type of the pp-wave spacetimes is \emph{N} or \emph{O} \cite{kramer,sg}. It is showed in \cite{hall1977} that the Petrov type is \emph{N} if the Weyl tensor is non-zero or conformally symmetric on $M$, and it is \emph{O} if the Weyl tensor vanishes. Therefore if the plane wave spacetime is Petrov type \emph{O}, then the functions $A(u),B(u)$ and $C(u)$ satisfy $A(u) = C(u)$ and $B(u) =0$.

The geodesic Lagrangian of the geodesic motion is given by
\begin{equation}
L = \frac{1}{2} g_{ab} \dot{x}^a \dot{x}^b - U(x^e),  \label{geodesic-lagr}
\end{equation}
where the dot represents the derivative with respect to the affine parameter $s$ and $U(x^e)$ is the potential function.
Using the pp-wave spacetime (\ref{pp-wave}), the geodesic Lagrangian (\ref{geodesic-lagr}) for pp-wave spacetimes in standard coordinates $x^a \equiv (u,v,y,z)$ takes such a form
\begin{equation}
L =  \frac{1}{2} \left( \dot{y}^2 +  \dot{z}^2 \right) - H(u,y,z) \dot{u}^2 - \dot{u} \dot{v} - U(u,v,y,z). \label{lagr}
\end{equation}
Considering this Lagrangian, one may obtain the geodesic equations of motion for the pp-wave spacetime by varying of the Lagrangian (\ref{lagr}) with respect to the coordinates $u, v, y$ and $z$ as follows:
\begin{eqnarray} && \ddot{u} - U_{,v} = 0, \label{lie1}
\\ && \ddot{v} + H_{,u} \dot{u}^2 + 2 H_{,y} \dot{y} \dot{u} + 2 H_{,z} \dot{u} \dot{z} - U_{,u} + 2 H U_{,v} =0, \label{lie2}
\\ &&  \ddot{y} + H_{,y} \dot{u}^2 + U_{,y} = 0, \label{lie3} \\& & \ddot{z} + H_{,z} \dot{u}^2 + U_{,z} = 0.  \label{lie4}
\end{eqnarray}
The \emph{energy functional} or \emph{Hamiltonian of the dynamical
system}, $E_{L}$, associated with the Lagrangian (\ref{lagr}) is
found as
\begin{eqnarray}
E_{L} &=&  \dot{x}^a p_a    - L \nonumber \\
&=& \frac{1}{2} \left( p_y^2  +  p_z^2 \right) + H(u,y,z) p_v^2 - p_{u} p_{v} + U(u,v,y,z), \label{E-L}
\end{eqnarray}
where $p_a = \frac{\partial L}{\partial \dot{x}^a}$, i.e. $p_u = -2 H \dot{u} - \dot{v}, \, p_v = - \dot{u}, \, p_y = \dot{y}$ and $p_z = \dot{z}$.

This study is organized as follows. In the following section, we present a detailed analysis of NGSs according to the isometry classes for pp-wave spacetimes. In section \ref{planewave}, we find the NGSs for plane wave spacetimes. In the last section, we conclude with a brief summary and discussions.

\section{Noether gauge symmetries for the pp-wave spacetimes}
\label{NGS}

The vector field ${\bf Y} = \xi \p_s + \eta^a \frac{\p}{\p x^a}$ is called a Noether gauge symmetry (NGS) of a Lagrangian $L(s,x^a,\dot{x^a})$ if there exists a gauge
function, $f(s,x^a)$, such that the Noether symmetry condition holds \cite{stephani,ibrahim}
\begin{equation}\label{noether1}
{\bf Y}^{[1]}L + L \, (D_{s}\xi) = D_{s} f ,
\end{equation}
where ${\bf Y^{[1]}} = {\bf Y}+\eta^a_{s} \frac{\partial}{\partial
\dot{x^a}}$ is the first prolongation operator, ${\eta^a_{s}} = D_{s}\eta^a - \dot{x^a} D_{s}\xi$, and ${D_{s}} =  \frac{\partial}{\partial s} + \dot{x}^a
\frac{\partial}{\partial x^a}$ is the total derivative operator.
The Noether gauge symmetries ${\bf Y}=\xi \p_s+\eta^1 \p_u+\eta^2 \p_v+ \eta^3 \p_y + \eta^4 \p_z$ of the geodesic Lagrangian (\ref{lagr}) are obtained from (\ref{noether1}), which explicitly
take the alternative form \cite{tsamparlis1,tsamparlis2}
\begin{eqnarray}
& & \xi_{,a} = 0, \label{neq-1} \\& & g_{ab} \eta^b_{,s} = f_{,a} \label{neq-2} \\& & \pounds_{\bf \eta} g_{ab} = \xi_{,s} g_{ab}  \label{neq-3} \\& &  \pounds_{\bf \eta} U = - \xi_{,s} U - f_{,s}  \label{neq-4}
\end{eqnarray}
where $\pounds_{\bf \eta}$ is the Lie derivative operator along ${\bf \eta} = \eta^1 \p_u+\eta^2 \p_v+\eta^3 \p_y+ \eta^4 \p_z$.
If the vector field ${\bf Y}$ is the NGS corresponding to the Lagrangian $L(s, x^a, \dot{x}^a)$, then
\begin{equation}\label{frstI}
{I} = \xi L + \left(\eta^a-\xi \dot{x}^a \right) \frac{\partial
L}{\partial \dot{x}^a} - f
\end{equation}
is the first integral associated with ${\bf Y}$. Then it follows from this relation for the geodesic Lagrangian (\ref{lagr}) that
\begin{equation}\label{frstI-2}
{I} = - \xi E_L -(2 H \eta^1 + \eta^2 ) \dot{u} - \eta^1 \dot{v} + \eta^3 \dot{y} + \eta^4 \dot{z} - f,
\end{equation}
where $E_L$ is given in (\ref{E-L}).

\subsection{\bf Noether gauge symmetry in standard coordinates}
\label{sec2-ngs}

For the metric (\ref{pp-wave}) in standard coordinates $x^a = (u,v,y,z)$, it follows from Eq. (\ref{neq-1}) that $\xi = \xi (s)$, and the remaining Noether equations (\ref{neq-2})-(\ref{neq-4}) lead to explicit form
\begin{eqnarray}
& & \eta^1_{,v} = 0, \quad \eta^1_{,y} - \eta^3_{,v} = 0, \quad \eta^1_{,z} - \eta^4_{,v} = 0, \label{neq1234} \\& & \eta^3_{,z} + \eta^4_{,y} = 0, \quad 2 \eta^3_{,y} - \xi_{,s} = 0, \quad 2 \eta^4_{,z} - \xi_{,s} = 0,  \label{neq567} \\& & \eta^2_{,v} + \eta^1_{,u} -\xi_{,s} = 0, \quad \eta^2_{,y} - \eta^3_{,u} + 2 H \eta^1_{,y} = 0, \quad  2 H \eta^1_{,z} + \eta^2_{,z} - \eta^4_{,u} = 0,  \label{neq89} \\& & \eta^1_{,s} + f_{,v} =0, \,\, \eta^3_{,s} - f_{,y} =0, \quad  \eta^4_{,s} - f_{,z} =0, \,\, \eta^2_{,s} + f_{,u} - 2 H f_{,v}  =0, \label{neq10-13}\\& & \eta^2_{,u} + 2 H \eta^1_{,u} + H_{,u} \eta^1 + H_{,y} \eta^3 + H_{,z} \eta^4 -H \xi_{,s} = 0, \label{neq14} \\& & U_{,u} \eta^1 + U_{,v} \eta^2 + U_{,y} \eta^3 + U_{,z} \eta^4 + U \xi_{,s} + f_{,s} = 0. \label{neq15}
\end{eqnarray}
On the one hand, for given $H(u,y,z)$, the Noether equations (\ref{neq1234})-(\ref{neq14}) determine the NGS generator ${\bf Y}$ of the Noether symmetry group. On the other hand, after finding the components $\xi, \eta^1, \eta^2, \eta^3$ and $\eta^4$ of the NGS generator ${\bf Y}$, it remains Eq. (\ref{neq15}) to be satisfied for the potential $U(u,v,y,z)$. In the following we solve these equations when the potential $U(u,v,y,z)$ vanishes, which gives a Noether symmetry called \emph{variational symmetry} \cite{stephani}. We find a solution (see the Appendix) for the NGS of a non-flat pp-wave spacetime as
\begin{eqnarray}
& & \xi = c_1 s + c_2, \quad \eta^1 = c_6 u + c_7, \label{xieta1}\\& & \eta^2 = -s f_{1,u} + y f_{2,u} + z f_{3,u} + \left( c_1 - c_6 \right) v + f_{4},  \label{eta2} \\& & \eta^3 = c_3 s + c_1 \frac{y}{2} + c_5 z + f_2, \quad \eta^4 = c_4 s -c_5 y +  c_1 \frac{z}{2} + f_3, \label{eta34} \\& & f = c_3 y + c_4 z + f_1, \label{f}
\end{eqnarray}
where $c_i$ are constant parameters, $i=1,...,7$ and $f_k, (_{k=1,...,4})$, are functions of $u$, and the following conditions have to be satisfied
\begin{eqnarray}
& &  f_{1,uu} -\left( c_3 H_{,y} + c_4 H_{,z} \right) =0, \label{ceq1} \\& & y f_{2,uu} + z f_{3,uu} + H_{,y} f_{2}+ H_{,z} f_{3} + \frac{c_1}{2} \left( y H_{,y}+ z H_{,z} - 2 H \right) \nonumber \\& & \qquad  + f_{4,u} + c_5 \left( z H_{,y} - y H_{,z} \right) + c_6 (u H_{,u} + 2 H ) + c_7 H_{,u} = 0. \label{ceq2}
\end{eqnarray}
Once $H(u,y,z)$ has been chosen, the differential equations (\ref{ceq1}) and (\ref{ceq2}) governs their behaviour. Fixing the function $H$ the special or non-special NGS admitted by a pp-wave spacetime can be deduced from Eqs. (\ref{ceq1}) and (\ref{ceq2}). For any form of $H$ one can have at most three independent non-special NGS.

From Eqs. (\ref{ceq1}) and (\ref{ceq2}) it seems that the functions $f_1 (u), f_4 (u)$ and the parameter $c_2$ are not fix the function $H$. Therefore, if the functions $f_1 (u)$ and $f_4 (u)$ are nonzero in the constraint equations (\ref{ceq1}) and (\ref{ceq2}), these constraint equations for any $H$ yield
\begin{eqnarray}
& &  f_{1,uu}  =0, \qquad f_{4,u}  = 0 \label{ceq1-1}
\end{eqnarray}
which give $f_1 = a_1 u +a_2$ and $f_4 =a_3$, where $a_1,a_2,a_3$ are constant parameters. Thus the components of NGS vector field ${\bf Y}$ and the gauge function $f$ become $\xi = c_2, \eta^1 = 0, \eta^2 = - a_1 s + a_3, \eta^3 = 0, \eta^4 = 0,$ and $f= a_1 u + a_2$. Here the constant parameter $a_2$ can be assumed to be zero without loss of generality.  Thus, a type N pp-wave spacetime can admit at most three non-special NGS such that
\begin{eqnarray}
& & {\bf k} = \partial_v, \qquad {\bf Y}_1 = \partial_s, \label{vf12} \\& & {\bf Y}_2 = -s \partial_v \quad {\rm with \, gauge\, term} \ f = u. \label{vf3}
\end{eqnarray}
From Eq.(\ref{frstI-2}) one can easily find the first integrals of ${\bf k}, {\bf Y}_1$ and ${\bf Y}_2$ as
\begin{eqnarray}
& & I_1 = -\dot{u}, \qquad I_2 = - E_L, \qquad I_3 = s \dot{u} -u, \label{fint123}
\end{eqnarray}
which means
\begin{eqnarray}
& &  u = -(I_1 s + I_3), \label{u2} \\
& & E_L = -I_2 \Leftrightarrow  \dot{y}^2 + \dot{z}^2 + 2 I_1 \left( \dot{v} - I_1 H \right) + 2 I_2 = 0, \label{E-L2}
\end{eqnarray}
where $I_1, I_2$ and $I_3$ are the Noether constants. It is well known that the normalizing condition of geodesics is
\begin{equation}
g_{ab} \dot{x}^a \dot{x}^b = -\kappa,
\end{equation}
where $\kappa = -1,0,+1$ for spacelike, null (lightlike) or timelike geodesics, respectively. This condition for pp-wave spacetime yields
\begin{equation}\label{tns}
-2H \dot{u}^2 -2 \dot{u} \dot{v} + \dot{y}^2 + \dot{z}^2 = -\kappa.
\end{equation}
For massive particles the curve parameter $s$ represents its proper time. In the case of light beam we use the curve parameter as an affine parameter $\lambda$ which has no physical meaning.
Using (\ref{u2}) in (\ref{tns}) and comparing this condition with (\ref{E-L2}) it is seen that $\kappa= 2 I_2$.
Additional NGSs make it possible to get the solution of geodesic equations or reduce the geodesic equations to the most convenient form in the associated isometry class. We conclude that the analytic solutions of geodesic equations can be found in some special cases.

In such a case one might ask if there can exist any further independent special or non-special NGS. The answer is in the positive for some form of $H$. We present some examples of pp-wave spacetimes admitting special NGS(s) in standard coordinates and give most of the results in Table \ref{T7}.

\subsubsection{Isometry class 1.}

In this class there exists only the KV ${\bf k}$ \cite{sg}. For this class we found two NGS ${\bf Y}_1$ and ${\bf Y}_2$ in addition to one KV ${\bf k}$, which means that this general spacetime admits minimal NGS algebra $\mathcal{N}_3 \supset \mathcal{G}_1$.
\begin{table}[!ht]
  \centering
  \caption{ NGSs, gauge functions and first integrals for the classes $1$, $1i$, $3$, $4$, $8$, $9$ and $Biv$. The vector field ${\bf Z}= 2u \p_u + y \p_y + z \p_z$ in class $Biv$ is a proper \emph{HKV}. \label{T7} }
  \resizebox{17cm}{!} {
\begin{tabular}{llll}
\hline\hline
 \,\,\, Class & \qquad NGS & $ f $ & First Integrals \\
 (subcase) &  &  &  \\
\hline
   1 & ${\bf k} = \partial_v$, \quad ${\bf Y}_1 = \partial_s$, \quad ${\bf Y}_2 = - s \partial_v$ & $0$, \quad $0$, \quad $u$ & $I_{1} = - \dot{u}$, \quad $I_{2} = - E_{L}$, \quad $I_{3} = s \dot{u} - u$ \\
     & & & \\

   1i & ${\bf k} = \partial_v$, \quad ${\bf Y}_1 = \partial_s$, \quad ${\bf Y}_2 = - s \partial_v$ & $0$, \quad $0$, \quad $u$ & $I_{1} = - \dot{u}$, \quad $I_{2} = - E_{L}$, \quad $I_{3} = s \dot{u} - u$ \\

    & ${\bf X}_2 = \partial_y$, \quad ${\bf X}_3 = u \partial_y + y \partial_v$, \quad ${\bf Y}_3 = s \partial_y$  & $0$, \quad $0$, \quad $y$ & $I_{4} = - \dot{y}$, \quad $I_{5} = u \dot{y} - y \dot{u}$, \quad $ I_{6} = s \dot{y} - y$  \\

\hline

   3 & ${\bf k} = \partial_v$, \quad ${\bf Y}_1 = \partial_s$, \quad ${\bf Y}_2 = - s \partial_v$ & $0$, \quad $0$, \quad $u$ & $I_{1} = - \dot{u}$, \quad $I_{2} = - E_{L}$, \quad $I_{3} = s \dot{u} - u$ \\
  & ${\bf X}_2 = \epsilon ( z \partial_y - y \partial_z ) + u \partial_u - v \partial_v \label{ic3-kv2}$ & $0$ & $I_{4} = - (2Hu -v)\dot{u} - u\dot{v} + \epsilon(z\dot{y} - y\dot{z})$ \\

\hline
  4 & ${\bf k} = \partial_v$, \quad ${\bf Y}_1 = \partial_s$, \quad ${\bf Y}_2 = - s \partial_v$ & $0$, \quad $0$, \quad $u$ & $I_{1} = - \dot{u}$, \quad $I_{2} = - E_{L}$, \quad $I_{3} = s \dot{u} - u$ \\
    & ${\bf X}_2 = \epsilon ( z \partial_y - y \partial_z ) + \partial_u \label{ic4-kv2}$ & $0$ & $I_{4} = - 2H \dot{u} - \dot{v} + \epsilon(z\dot{y}-y\dot{z})$ \\

\hline

  8 & ${\bf k} = \partial_v$, \quad ${\bf Y}_1 = \partial_s$, \quad ${\bf Y}_2 = - s \partial_v$ & $0$, \quad $0$, \quad $u$ & $I_{1} = - \dot{u}$, \quad $I_{2} = - E_{L}$, \quad $I_{3} = s \dot{u} - u$ \\

    & ${\bf X}_2 = \partial_u$, & $0$, & $I_{4} = -2 H \dot{u}-\dot{v}$, \\
    & ${\bf X}_3 = \epsilon (u \p_u -v \p_v) + \eta \p_y + \sigma \p_z$ & $0$ & $I_{5} = \epsilon (-2Hu\dot{u}-u\dot{v}+v\dot{u})+\eta \dot{y}+\sigma \dot{z}$ \\

     & & & \\

 8($\epsilon =0$) & ${\bf k} = \partial_v$, \quad ${\bf Y}_1 = \partial_s$, \quad ${\bf Y}_2 = - s \partial_v$ & $0$, \quad $0$, \quad $u$ & $I_{1} = - \dot{u}$, \quad $I_{2} = - E_{L}$, \quad $I_{3} = s \dot{u} - u$ \\

   & ${\bf X}_2 = \partial_u$, \quad ${\bf X}_3 = \eta \p_y + \sigma \p_z$ & $0$, \quad $0$ & $I_{4} = -2 H \dot{u}-\dot{v}$, \quad $I_{5} = \eta\dot{y}+\sigma\dot{z}$ \\
   & ${\bf X}_4 = \eta (y \p_v + u \p_y ) + \sigma (z \p_v + u \p_z )$,  &  $0$, & $I_{6} = \eta(-y\dot{u}+u\dot{y})+\sigma(-z\dot{u}+u\dot{z})$, \\
   & ${\bf Y}_3 = s \left( \eta \p_y + \sigma \p_z \right)$ & $\eta y + \sigma z$ & $I_{7} = s(\eta\dot{y}+\sigma\dot{z}) - (\eta y + \sigma z)$ \\

\hline

  9 & ${\bf k} = \partial_v$, \quad ${\bf Y}_1 = \partial_s$, \quad ${\bf Y}_2 = - s \partial_v$ & $0$, \quad $0$, \quad $u$ & $I_{1} = - \dot{u}$, \quad $I_{2} = - E_{L}$, \quad $I_{3} = s \dot{u} - u$ \\

   & ${\bf X}_2 = \partial_u$, \quad ${\bf X}_3 = \p_z + \eta ( u \partial_u - v \partial_v)$ & $0$, \quad $0$ & $I_{4} = -2 H \dot{u}-\dot{v}$, \quad $I_{5} = \dot{z} - \eta u (2 H \dot{u} + \dot{v}) + \eta v \dot{u}$ \\

   & ${\bf X}_4 = \partial_y - \sigma (u \partial_u - v \partial_v)$ & $0$ & $I_{6} = \dot{y} + \sigma u (2 H \dot{u} + \dot{v}) - \sigma v \dot{u}$\\

   & ${\bf X}_5 = u (\eta \partial_y + \sigma \partial_z) + (\eta y + \sigma z) \partial_v$ & $0$ & $I_{7} = u (\sigma \dot{z} + \eta \dot{y}) - (\eta y + \sigma z) \dot{u}$ \\

   & ${\bf Y}_3 = s (\eta \partial_y + \sigma \partial_z)$ & $\eta y + \sigma z$ & $I_{8} = \eta (s \dot{y} - y) + \sigma (s \dot{z} - z)$ \\
\hline

 Biv & ${\bf k} = \partial_v$, \quad ${\bf Y}_1 = \partial_s$, \quad ${\bf Y}_2 = - s \partial_v$ & $0$, \quad $0$, \quad $u$ & $I_{1} = - \dot{u}$, \quad $I_{2} = - E_{L}$, \quad $I_{3} = s \dot{u} - u$ \\

     & ${\bf X}_2 = \partial_u$, \quad  ${\bf X}_3 = \sigma \p_y + \rho \p_z$ & $0$, \quad $0$ & $I_{4} = -2 H \dot{u}-\dot{v}$, \quad $I_{5} = \sigma \dot{y} + \rho \dot{z}$ \\

  &  ${\bf X}_4 = (\sigma y + \rho z) \p_v + u (\sigma \p_y + \rho \p_z)$  & $0$ & $I_{6} = -(\sigma y + \rho z)\dot{u} + \sigma u \dot{y} + \rho u \dot{z}$\\

  & ${\bf Y}_3 = s (\sigma \p_y + \rho \p_z)$, & $\sigma y + \rho z$, & $I_{7} = \sigma (s \dot{y} - y) + \rho (s \dot{z} - z)$,  \\
  & ${\bf Y}_4 = 2s \p_s + {\bf Z}$ & $0$ & $I_{8} = -2 s E_{L} - 2 u (2 H \dot{u} + \dot{v}) + y \dot{y} + z \dot{z}$ \\

\hline\hline
\end{tabular}
}
\end{table}

\subsubsection{Isometry class 1i.}

In this class $H$ is independent of one of the spatial coordinates, i.e. $H = H(u,z)$ \cite{kt}.  Then the conditions (\ref{ceq1}) and (\ref{ceq2}), and exclusion of type \emph{O} and plane wave spacetimes give $c_4 =0, c_5 = 0, f_{1,uu}=0, f_{2,uu} = 0$, i.e. $f_1 =a_1 u+ a_2, f_2 = a_3 u + a_4$, and
\begin{eqnarray}
& & \left( f_3 + c_1 \frac{z}{2} \right) H_{,z} + (c_6 u + c_7) H_{,u} +(2 c_6 -c_1) H + z f_{3,uu} + f_{4,u}=0, \label{ic1-1}
\end{eqnarray}
where $a_1, a_2, a_3$ and $a_4$ are constants. It is easily found that if $H(u,z)$ is an arbitrary function then it follows from (\ref{ic1-1}) that the NGSs which can occur are ${\bf k}, {\bf X}_2, {\bf X}_3, {\bf Y}_1, {\bf Y}_2$ and ${\bf Y}_3$. Then using the Table \ref{T7} and integrating, we find that
\begin{eqnarray}
& & u = -(I_1 s + I_3), \qquad y = -(I_4 s + I_6), \qquad I_5 = I_3 I_4 - I_1 I_6, \label{case1i-1}  \\& & \dot{z}^2 + 2 I_1 \left( \dot{v} - I_1 H \right) = - (\epsilon + I_4^2). \label{case1i-2}
\end{eqnarray}
Therefore, we have the exact form of $u$ and $y$, and it is seen that we need an extra equation to get $v$ and $z$.

For this class the function $H$ could also depend on the coordinates $u$ and $y$, i.e. $H=H(u,y)$. Therefore the conditions (\ref{ceq1}) and (\ref{ceq2}) yield $c_3 =0, c_5 = 0, f_{1,uu}=0, f_{3,uu} = 0$, i.e. $f_1 =a_1 u+ a_2, f_3 = a_3 u + a_4$, and
\begin{eqnarray}
& & \left( f_2 + c_1 \frac{y}{2} \right) H_{,y} + (c_6 u + c_7) H_{,u} +(2 c_6 -c_1) H + y f_{2,uu} + f_{4,u}=0, \label{ic1-2}
\end{eqnarray}
which gives rise to same NGSs appear in $H = H(u,z)$, where the NGSs, gauge functions and first integrals come with the coordinate $z$ instead of the coordinate $y$.

\subsubsection{Isometry class 3.}

In this class, the differential equation for $H$ \cite{sg} is
\begin{equation}
u H_{,u} + 2 H + \epsilon ( z H_{,y} - y H_{,z} ) =0, \label{ic3-deq}
\end{equation}
which give the form $H = u^{-2} W(\mu,\nu)$, where $\mu = z \sin \phi - y \cos \phi, \, \nu = y \cos\phi + z \sin\phi, \phi = \epsilon \ln |u|$ and $\epsilon$ is an arbitrary constant. Here we used the same coordinate transformation as with Sippel and Goenner \cite{sg}. Keane and Tupper \cite{kt} have considered the coordinate transformation $s = y \sin \phi - z \cos \phi, \, t = y \cos\phi + z \sin\phi, \phi = \epsilon \ln |u|$, but this yields $H = u^{-1} W(s,t)$. From the condition (\ref{ceq2}) and the differential equation (\ref{ic3-deq}) for $H$, we observe that $c_5 = \epsilon k_1$ and $c_6 = k_1$, where $k_1$ is a constant parameter. Using the considered coordinate transformation above, the conditions (\ref{ceq1}) and (\ref{ceq2}) become
\begin{eqnarray}
& & u^2 f_{1,uu} + \left[ (c_3 \cos\phi - c_4 \sin\phi )W_{,\mu} - (c_3 \sin\phi + c_4 \cos\phi )  W_{,\nu} \right] = 0, \label{ic3-1} \\& & u^2 \left[ (\nu \sin\phi - \mu \cos\phi) f_{2,uu} + (\mu \sin\phi + \nu \cos\phi ) f_{3,uu} + f_{4,u} \right] \nonumber \\& &  \quad  + (-f_2 \cos\phi + f_3 \sin\phi ) W_{,\mu} + (f_2 \sin\phi + f_3 \cos\phi ) W_{,\nu} \nonumber \\& &  \quad   + \frac{c_1}{2} \left( \mu W_{,\mu} + \nu W_{,\nu} -2 W \right) + c_7 \frac{\epsilon}{u} \left( \nu W_{,\mu} - \mu W_{,\nu} -\frac{2}{\epsilon } W \right)  =0. \label{ic3-2}
\end{eqnarray}
Then it follows that if $W$ is an arbitrary function, the NGSs can occur as two KVs ${\bf k}, {\bf X}_2, {\bf Y}_1$ and ${\bf Y}_2$. Thus, for the geodesic equations in this case, it follows from the Table \ref{T7} that $u= -(I_1 s + I_3)$ and
\begin{eqnarray}
& & \dot{y}^2 + \dot{z}^2 + 2 I_1 (\dot{v} - I_1 H) + \kappa = 0, \label{case3-1}\\& & \epsilon ( z \dot{y} + y \dot{z} ) + (I_1 s + I_3) (\dot{v} - 2 I_1 H ) -I_1 v -I_4 =0, \label{case3-2}
\end{eqnarray}
where $\kappa = 2 I_2 =0, \pm 1$. When $W$ is an arbitrary function, this result represents that there are three unknowns $\dot{y}, \dot{z}$ and $\dot{v}$, but two differential equations given above. Therefore, we need at least one additional differential equation to solve the geodesic equations completely, if it is possible to solve the obtained system of differential equations.

It is too difficult to solve the above partial differential equations (\ref{ic3-1}) and (\ref{ic3-2}) in general but we can give some special solutions. For example, assuming $f_2= - k_2 \cos\phi$ and $f_3 = k_2 \sin\phi$, where $k_2$ is a constant parameter, we find from Eqs. (\ref{ic3-1}) and (\ref{ic3-2}) that for $W=\epsilon \mu \left( \nu + \epsilon \mu /2 \right) + K(\nu)$, $K(\nu)$ is a function of integration, there is an additional KV as
\begin{equation}
{\bf X}_3 = \frac{\epsilon}{u} ( y \sin\phi + z \cos\phi ) \partial_v - \cos\phi \partial_y + \sin\phi \partial_z,  \label{ic3-kv3}
\end{equation}
with $I_{5} = - \frac{\epsilon}{u}(y sin\phi + z cos\phi) \dot{u} - \cos\phi \dot{y} + sin\phi \dot{z}$ which means that the spacetime admits $\mathcal{N}_5 \supset \mathcal{G}_3$.  If $W (\mu,\nu)$ satisfies the condition
\begin{equation}
\mu W_{,\mu} + \nu W_{,\nu} - 2 W  =0, \label{ic3-deq2}
\end{equation}
which indicates that the constant parameter $c_1$ need not to be zero, then we have a special NGS
\begin{equation}
{\bf Y}_3 = 2 s \partial_s + {\bf Z}, \label{ic3-ngs}
\end{equation}
with the first integral $I_{6} = -2sE_L - 2v \dot{u} + y \dot{y} + z \dot{z},$ where ${\bf Z} = 2 v \partial_v + y \partial_y + z \partial_z$ is a HKV. This means that the spacetime admits $\mathcal{N}_6 \supset \mathcal{H}_4 \supset \mathcal{G}_3$. Note that using $W=\epsilon \mu \left( \nu + \epsilon \mu /2 \right) + K(\nu)$ in Eq.(\ref{ic3-deq2}) yields $K(\nu) = K_0 \nu^2$, where $K_0$ is a constant of integration. Also it is noted that even we have two more constants of motion $I_5$ and $I_6$ which give two additional differential equations, the related geodesic equations could not be solved yet.

\subsubsection{Isometry class 4.}

The differential equation for $H$ for this class \cite{sg} is
\begin{equation}
H_{,u} + \epsilon ( z H_{,y} - y H_{,z} ) =0, \label{ic4-deq}
\end{equation}
which gives $H= W(\mu,\nu)$, where $\phi = \epsilon u$, $\epsilon$ is an arbitrary constant. From the condition (\ref{ceq2}) and the differential equation (\ref{ic4-deq}) for $H$, we observe that $c_5 = \epsilon k_1$ and $c_7 = k_1$, where $k_1$ is a constant parameter. Furthermore, the constraint conditions (\ref{ceq1}) and (\ref{ceq2}) of this class yield
\begin{eqnarray}
& & f_{1,uu} + \left[ (c_3 \cos\phi - c_4 \sin\phi )W_{,\mu} - (c_3 \sin\phi + c_4 \cos\phi )  W_{,\nu} \right] = 0, \label{ic4-1} \\& &  (\nu \sin\phi - \mu \cos\phi) f_{2,uu} + (\mu \sin\phi + \nu \cos\phi ) f_{3,uu} + f_{4,u} \nonumber \\& &  \quad + (-f_2 \cos\phi + f_3 \sin\phi ) W_{,\mu} + (f_2 \sin\phi + f_3 \cos\phi ) W_{,\nu} \nonumber \\& &  \quad + \frac{c_1}{2} \left( \mu W_{,\mu} + \nu W_{,\nu} -2 W \right) + c_6 \left[ \phi \left( \nu W_{,\mu} - \mu W_{,\nu}\right) + 2 W \right]  =0. \label{ic4-2}
\end{eqnarray}
It is also too difficult to get a solution of the above constraint equations in general. After some algebra we find from the above constraint equations that $f_1= a_1 u + a_2, \, f_4 = a_3$ and $c_3 = c_4 = c_6 = 0$. Then it follows that if $W$ is an arbitrary function, the NGSs are two non-special NGSs ${\bf Y}_1$ and ${\bf Y}_2$, and two KVs ${\bf k}$ and ${\bf X}_2$.
For this class, the geodesic equations from the Table \ref{T7} gives
\begin{eqnarray}
& & u= - (I_1 s + I_3), \\& & \dot{y}^2 + \dot{z}^2 + 2 I_1 (\dot{v} - I_1 H) + \kappa = 0, \label{case4-1}\\& & \epsilon ( z \dot{y} - y \dot{z} ) -\dot{v} + 2 I_1 H -I_4 =0, \label{case4-2}
\end{eqnarray}
which are also under-determined system of ordinary differential equations for $\dot{y}, \dot{z}$ and $\dot{v}$. Therefore, it is also not possible to solve these kind of system just like the isometry class 3.

Under the assumption $f_2= - k_2 \cos\phi$ and $f_3 = k_2 \sin\phi$, one can obtain an additional KV as
\begin{equation}
{\bf X}_3 = \epsilon  ( y \sin\phi + z \cos\phi ) \partial_v - \cos\phi \partial_y + \sin\phi \partial_z,  \label{ic4-kv3}
\end{equation}
with $I_{5} = -\epsilon(y sin\phi + z cos\phi) \dot{u} - cos\phi \dot{y} + sin\phi \dot{z}$ for $W=\mu^2 /2 + L(\nu)$, where $L(\nu)$ is an integration function. In this case if $W$ satisfies the condition (\ref{ic3-deq2}) we find the same NGS as (\ref{ic3-ngs}). Thus the spacetime admits $\mathcal{N}_6 \supset \mathcal{H}_4 \supset \mathcal{G}_3$.

\subsubsection{Isometry class 8.}

In this case, the conditions of \cite{sg} on $H$ are $H_{,u} = 0$ which means $H= H(y,z)$, and
\begin{equation}
\eta H_{,y} + \sigma H_{,z} + 2 \epsilon H = 0, \label{ic8}
\end{equation}
where $\eta, \sigma$ and $\epsilon (\neq 0)$ are constants such that $\eta^2 + \sigma^2 \neq 0$. We will partially use Keane and Tupper's \cite{kt} notation in which $t = \eta z -\sigma y, w = \eta y + \sigma z$ and $\delta = -\epsilon/(\eta^2 + \sigma^2)$, where we have not considered $s$ instead of $t$ because of that $s$ is arc length parameter throughout this study. Thus the condition (\ref{ic8}) leads to the form $H=W(t) exp(2 \delta w)$. If $W(t)$ is an arbitrary function then it follows from the conditions (\ref{ceq1}) and (\ref{ceq2}) that $c_1$ and $c_5$ vanish, and
\begin{eqnarray}
& & f_{1,uu} =0, \qquad f_{2,uu} =0, \qquad f_{3,uu} =0, \qquad f_{4,u} =0, \label{ic8-1} \\& &
(\eta f_3 - \sigma f_2 ) W'(t) + 2 \left[ \delta (\eta f_2 + \sigma f_3) + c_6 \right] W(t) = 0. \label{ic8-2}
\end{eqnarray}
From the last Eq.(\ref{ic8-2}) we find $f_2 = \eta, f_3 = \sigma$ and $c_6 = \epsilon$ for arbitrary $W(t)$. These results together with (\ref{ic8-1})  imply that we get the three KVs ${\bf k}, {\bf X}_2, {\bf X}_3$ and the two NGSs ${\bf Y}_1$ and ${\bf Y}_2$. In this class the spacetime admits $\mathcal{N}_5 \supset \mathcal{G}_3$.
Furthermore we can find a specialization for $\epsilon \neq 0$ only if $W(t)=K exp(ct)$, where $K$ and $c$ are non-zero constants, but this corresponds to isometry class 9.

For this class, the first integrals given in Table \ref{T7} yields
\begin{eqnarray}
& & u = -( I_1 s + I_3), \qquad v = 2 I_1 \int{H ds} - I_4 s + v_0, \\& & \eta \dot{y} + \sigma \dot{z} = \epsilon I_1 v + \epsilon I_4 ( I_1 s + I_3) + I_5,  \label{geq1-8-1} \\& & \dot{y}^2 + \dot{z}^2 + 2 I_1 (I_1 H - I_4) + \kappa =0, \label{geq1-8-2}
\end{eqnarray}
where $\kappa = 2 I_2$, and $v_0$ is an integration constant. Here we could not find the solution of the above geodesic equations (\ref{geq1-8-1}) and (\ref{geq1-8-2}) explicitly.

\subsubsection{Isometry class 8($\epsilon =0$).}

In this case the condition (\ref{ic8}) yields that the function $H$ has the form $H = W(t)$.  For this class, the conditions (\ref{ceq1}) and (\ref{ceq2}) lead to $c_5 = 0, c_3 = k_1 \eta, c_4=k_2 \sigma$ and the following equations
\begin{eqnarray}
& & f_{1,uu} = 0, \qquad f_{2,uu} =0, \qquad f_{3,uu} =0, \label{ic8-2-1} \\& & \left( c_1 \frac{t}{2} + \eta f_3 - \sigma f_2 \right) W'(t) + (2 c_6 -c_1) W(t) + f_{4,u} =0, \label{ic8-2-2}
\end{eqnarray}
where $k_1$ and $k_2$ are constants. Eq.(\ref{ic8-2-1}) implies $f_1 = a_1 u + a_2, f_2 = a_3 u + a_4, f_3 = a_5 u + a_6$ where $a_1,...,a_6$ are constant parameters. Then differentiating Eq.(\ref{ic8-2-2}) with respect to $u$ it reduces to
\begin{equation}
(a_5 \eta - a_3 \sigma) W'(t) + f_{4,uu} = 0, \label{ic8-2-3}
\end{equation}
which gives $a_3 = k_2 \eta, a_5 = k_2 \sigma$ and $f_4 = a_7 u + a_8$ for arbitrary $W(t)$, where $k_2, a_7$ and $a_8$ are constants. Thus Eq.(\ref{ic8-2-2}) implies
\begin{equation}
\left( c_1 \frac{t}{2} + a_6 \eta - a_4 \sigma \right) W'(t) + (2 c_6 -c_1) W(t) + a_7 =0. \label{ic8-2-4}
\end{equation}
Therefore if $W(t)$ is arbitrary then we have four KVs ${\bf k}, {\bf X}_2, {\bf X}_3, {\bf X}_4$ and three NGSs ${\bf Y}_1, {\bf Y}_2, {\bf Y}_3$. Thus the spacetime with $H=W(t)$ admits $\mathcal{N}_7 \supset \mathcal{G}_4$. For this class, using the Table \ref{T7}, we obtain the following relations
\begin{eqnarray}
& & u = -( I_1 s + I_3), \qquad v = 2 I_1 \int{W (t) ds} - I_4 s + v_0, \label{8e-geq-1}  \\& & \eta y + \sigma z = I_5 s - I_7,  \qquad I_6 = - (I_1 I_7 + I_3 I_5), \label{8e-geq-2} \\& & \dot{y}^2 + \dot{z}^2 + 2 I_1 (I_1 W - I_4) + \kappa =0, \label{8e-geq-3}
\end{eqnarray}
where $\kappa = 2 I_2$, $v_0$ is an integration constant, and $t= \eta z - \sigma y$. It is not  possible to solve exactly the geodesic equation (\ref{8e-geq-3}) due to the arbitrariness of  $W(t)$.

The Eq.(\ref{ic8-2-4}) gives rise to some specific examples of $W(t)$. The restrictions $c_6 = c_1, a_7 =0, a_4 = k_3 \eta$ and $a_6 = k_3 \sigma$ in (\ref{ic8-2-4}) leads to $W(t) = K t^{-2}$ with
algebra $\mathcal{N}_8 \supset \mathcal{H}_5 \supset \mathcal{G}_4$, which is class \emph{Biv}. If $a_7 = 0, a_4 = k_3 \eta, a_6 = k_3 \sigma$ and $c_6 = \frac{(2-c)}{4} c_1$ then these restrictions in (\ref{ic8-2-4}) lead to $W(t) = K t^c$ where $c (\neq 2)$ is a constant. For this form of $W(t)$ we have an algebra consisting of KVs ${\bf k}, {\bf X}_2, {\bf X}_3, {\bf X}_4$ and NGSs ${\bf Y}_1, {\bf Y}_2, {\bf Y}_3$, and
\begin{equation}
{\bf Y}_4 = 2 s \p_s + {\bf Z},
\end{equation}
with the first integral $ I_{8} = -2 s E_{L} - \left[ 2H \frac{(2-c)}{2} u + \frac{(2+c)}{2}v \right] \dot{u}-\frac{(2-c)}{2}u\dot{v} + y \dot{y} + z \dot{z}$ where ${\bf Z}$ is the HKV in the form
\begin{equation}
{\bf Z} = \frac{(2-c)}{2} u \p_u + \frac{(2+c)}{2} v \p_v + y \p_y + z \p_z,
\end{equation}
therefore this spacetime also admits $\mathcal{N}_8 \supset \mathcal{H}_5 \supset \mathcal{G}_4$. Here, it is too difficult to solve the geodesic equation (\ref{8e-geq-3}) because of the form of $W(t)$.

\subsubsection{Isometry class 9.}

This case is a special case of isometry class 8, and $H$ has the form $H = K e^{2 (\sigma y - \eta z)}$. For this class we found \emph{five} KVs, i.e. ${\bf k}, {\bf X}_2, {\bf X}_3, {\bf X}_4, {\bf X}_5$ and \emph{three} NGSs ${\bf Y}_1, {\bf Y}_2, {\bf Y}_3$.
The spacetime admits the $\mathcal{N}_8 \supset \mathcal{G}_5$.

In this class, the first integrals from the Table \ref{T7} give
\begin{eqnarray}
& & u = -( I_1 s + I_3), \qquad v = 2 K I_1 \int{ e^{ 2(\sigma y -\eta z )} ds} - I_5 s + v_0, \label{9-geq-1}  \\& & \eta y + \sigma z  =  (\sigma I_6 + \eta I_7) s  - I_4,  \qquad I_8 = - I_3 (\sigma I_6 + \eta I_7)- I_1 I_4, \label{9-geq-2} \\& & \dot{y}^2 + \dot{z}^2 + 2 I_1 (I_1 W - I_4) + \kappa =0, \label{9-geq-3} \\& & \dot{y} = - \sigma I_1 v - \sigma I_5 (I_1 s + I_3) + I_7, \label{9-geq-4}   \\& & \dot{z} = \eta I_1 v + \eta I_5 (I_1 s + I_3) + I_6,  \label{9-geq-5}
\end{eqnarray}
where $I_1 \neq 0, \, \kappa = 2 I_2$ and $v_0$ is an integration constant. Then, in a special case of isometry class 9, where only non-zero Noether constants are $I_1$ and $I_5$, we found the analytic solution of geodesic equations for $H = K \exp(2 \ell y)$ of the form
\begin{eqnarray}
& & u (\lambda) = -I_1 \lambda,  \\& &  v (\lambda) = \frac{I_5}{2 \ell \sqrt{\beta}} \left[ 2 \tanh \ell \sqrt{\beta} (y_0 - \lambda) + \ln \left( \frac{ \tanh \ell \sqrt{\beta} (y_0 - \lambda) -1}{\tanh \ell \sqrt{\beta} (y_0 - \lambda) + 1} \right) \right] \nonumber \\& & \qquad \qquad - I_5 \lambda + v_0,  \\& & y(\lambda) = \frac{1}{2 \ell} \ln \left[ \frac{\ell \beta}{\alpha} \left(1 - \tanh^2 \ell \sqrt{\beta} (y_0 - \lambda) \right) \right], \\\ & & z(\lambda) = - \frac{\eta}{\sigma} y (\lambda),
\end{eqnarray}
where $ \sigma \neq 0,\, \ell = (\sigma^2 + \eta^2) / \sigma, \alpha = 2 \sigma K I_1^2$, $\beta= 2 \sigma I_1 I_5 / \ell, \sigma^2 + \eta^2 \neq 0$, $v_0$ and $y_0$ are integration constants. Since $I_2$ vanishes, this means $\kappa = 0$, i.e. this corresponds to the massless particles,  therefore we use the curve parameter $\lambda$ instead of $s$.

There may be exists other special cases of Noether constants whether some of these constants vanish or not. In other special cases, it is possible to find some analytical solutions of the geodesic equations, but this is a subject of another study.

\subsubsection{Class Biv.}

In this class the function $H$ is given by $H= l (\sigma z - \rho y)^{-2}$. For this class
the number of NGSs we found from the constraint conditions (\ref{ceq1}) and (\ref{ceq2}) is \emph{eight}: ${\bf k}, {\bf X}_2, {\bf X}_3, {\bf X}_4, {\bf Y}_1, {\bf Y}_2, {\bf Y}_3, {\bf Y}_4$. The spacetime admits the $\mathcal{N}_8 \supset \mathcal{H}_5 \supset \mathcal{G}_4$.

Now, for this class, it follows from the first integrals given in Table \ref{T7} that
\begin{eqnarray}
& & u = -( I_1 s + I_3), \qquad \eta y + \sigma z  =  I_7 s - I_4,  \label{b4-geq-1}  \\& & \dot{v} = 2 I_1 H  - I_6,  \qquad I_8 = - (I_1 I_4 + I_3 I_7), \label{b4-geq-2} \\& & \dot{y}^2 + \dot{z}^2 + 2 I_1 (I_1 H - I_6) + \kappa =0, \label{b4-geq-3} \\& & y \dot{y} + z \dot{z} - 2 (I_2 - I_1 I_6) s - 2 I_3 I_6 - I_5 = 0, \label{b4-geq-4}
\end{eqnarray}
where $\kappa = 2 I_2$. Here we consider only one special case of class Biv, when $I_1, I_2$ and $I_8$ are non-zero Noether constants. For this special case of class Biv, the general solution of geodesic equations are given by
\begin{eqnarray}
& & u (s) = -I_1 s,  \quad  v (s) =  \frac{2 l \, \rho I_1}{k \sqrt{k y_0 \kappa}} {\rm arctanh} \left( \frac{\rho \kappa s}{\sqrt{k y_0 \kappa}} \right)  +  v_0, \\& & y(s) = \mp \sqrt{y_0 - \frac{\kappa \rho^2}{k} s^2 }, \quad z(s) = - \frac{\sigma}{\rho} y (s),
\end{eqnarray}
where $k = \sigma^2 + \rho^2$, and $v_0$ and $y_0$ are constants of integration.

In this class, it is also possible other special cases of class Biv in which one can obtain some analytical solutions of geodesic equations, but this is again a subject of another study.

\subsection{\bf Noether gauge symmetry in polar coordinates}

Most of the Killing symmetries obtained in references \cite{tupper2003,kt,sg,mm} are in the polar coordinates. Therefore for the moment we will write down the metric (\ref{pp-wave}) in polar coordinates $ y = r \cos \theta$ and $z = r \sin \theta$ as
\begin{equation}
ds^2 = -2 H(u,r,\theta) du^2 -2 du dv + dr^2 + r^2 d\theta^2. \label{pp-wave2}
\end{equation}
Using this form of pp-wave spacetimes, a point-like Lagrangian density in polar coordinates takes
such a form
\begin{equation}
L =  \frac{1}{2} \left( \dot{r}^2 +  r^2 \dot{\theta}^2 \right) - H(u,r,\theta) \dot{u}^2 - \dot{u} \dot{v} - U(u,v,r,\theta). \label{lagr2}
\end{equation}
Thus the explicit form of Noether equations (\ref{neq-2})-(\ref{neq-4}) together with $\xi = \xi (s)$ from Eq. (\ref{neq-1}) becomes
\begin{eqnarray}
& & \eta^1_{,v} = 0, \quad \eta^1_{,r} - \eta^3_{,v} = 0, \quad \eta^1_{,\theta} - r^2 \eta^4_{,v} = 0, \quad \eta^3_{,\theta} + r^2 \eta^4_{,r} = 0, \label{pneq1234} \\& & 2 \eta^3_{,r} - \xi_{,s} = 0, \quad 2 \eta^4_{,\theta} + \frac{2}{r} \eta^3- \xi_{,s} = 0, \quad \eta^2_{,v} + \eta^1_{,u} -\xi_{,s} = 0, \label{pneq567} \\& & 2 H \eta^1_{,\theta} + \eta^2_{,\theta} - r^2 \eta^4_{,u} = 0, \quad 2 H \eta^1_{,r} + \eta^2_{,r} - \eta^3_{,u}  = 0,  \label{pneq89} \\& & \eta^1_{,s} + f_{,v} =0, \, \,  \eta^3_{,s} - f_{,r} =0, \,\, r^2 \eta^4_{,s} - f_{,\theta} =0, \,\, \eta^2_{,s} + f_{,u} - 2 H f_{,v} =0, \label{pneq10-13}\\& & \eta^2_{,u} + 2 H \eta^1_{,u} + H_{,u} \eta^1 + H_{,r} \eta^3 + H_{,\theta} \eta^4 -H \xi_{,s} = 0, \label{pneq14} \\& & U_{,u} \eta^1 + U_{,v} \eta^2 + U_{,r} \eta^3 + U_{,\theta} \eta^4 + U \xi_{,s} + f_{,s} = 0. \label{pneq15}
\end{eqnarray}
Thus a solution of the above NGS equations in polar coordinates, which can be obtained by a similar calculation given in the Appendix, yields
\begin{eqnarray}
& & \xi = c_1 s + c_2, \quad \eta^1 = c_6 u + c_7, \\& & \eta^2 = -s g_{1,u} + r \cos \theta g_{2,u} + r \sin \theta g_{3,u} + \left( c_1 - c_6 \right) v + g_{4},  \\& & \eta^3 = \left( c_4 s + g_2 \right) \cos \theta +  \left( c_3 s + g_3 \right) \sin \theta  + c_1 \frac{r}{2}, \\& & \eta^4 = \left( c_3 s + g_3 \right) \frac{\cos \theta}{r} - \left( c_4 s + g_2 \right) \frac{\sin \theta}{r} +  c_5 , \\& & f = r ( c_3 \sin \theta + c_4  \cos \theta )  + g_1
\end{eqnarray}
where $c_i$ are constant parameters, $i=1,...,7$ and $g_k, (_{k=1,...,4})$, are functions of $u$, and the following conditions have to be satisfied
\begin{eqnarray}
& & \quad g_{1,uu} - c_3 \left( \frac{\cos \theta}{r} H_{,\theta} -\sin \theta H_{,r} \right) -c_4 \left( \cos \theta H_{,r} + \frac{\sin \theta}{r} H_{,\theta} \right) = 0, \label{ceq1p} \\& &  \quad r \left( \cos \theta  g_{2,uu} + \sin \theta g_{3,uu} \right) + H_{,r} \left( \cos \theta g_2 + \sin \theta g_{3} \right) \nonumber \\& & \qquad + \frac{H_{,\theta}}{r} \left( \cos\theta g_3 - \sin\theta g_2 \right) + g_{4,u} + \frac{c_1}{2} \left( r H_{,r} - 2 H \right) \nonumber \\& & \qquad + c_5 H_{,\theta} + c_6 \left( u H_{,u} + 2 H \right) + c_7 H_{,u}  = 0. \label{ceq2p}
\end{eqnarray}
Here it is also seen that the functions $g_1 (u),g_4(u)$ and the parameter $c_2$ are not fix the function $H$. This implies again that we have three non-special NGSs given by (\ref{vf12}) and (\ref{vf3}).

Four possible classes of solutions depending on the form of metric function $H = \tau (u) r^2 + \delta (\theta) r^{-2}$ arise from the conformal Killing equations of pp-wave spacetime \cite{kt}
\begin{enumerate}
\item[{\it A} :] $\delta(\theta) = \ell e^{2 m \theta}$, where $\ell \neq 0, m \neq 0$ are constants.

\item[{\it B} :] $\delta(\theta) = \ell \left( \eta \sin \theta - \sigma \cos \theta \right)^{-2}$, where $\ell \neq 0, \eta$ and $\sigma$ are nonzero constants.

\item[{\it C} :] $\delta(\theta)$ is a constant.

\item[{\it D} :] $\delta(\theta)$ is arbitrary function.

\end{enumerate}
We consider the above classes {\it A-D} according to form of the function of $H$ given in \cite{kt}, and find NGSs of the geodesic Lagrangian for pp-wave spacetimes of those classes given in Table \ref{T1}, where the second column denotes the NGSs, the third and fourth columns denote gauge functions of NGSs and first integrals of geodesic equations correspondingly. For Class B; $ G_1 = u \sin( q / 2 u) $, $G_2 = u \cos( q / 2 u) $, $ D_1 = \sin(\sqrt{2 q} u) $, $D_2 = \cos(\sqrt{2 q} u) $, $P_1(\theta) = \sigma \cos\theta + \rho \sin\theta$, $P_2(\theta) = \sigma \sin\theta - \rho \cos\theta $.  We have seen that the classes A and D belong to isometry class 1.

If the function $\tau(u)$ is chosen zero, we get classes type iv \cite{kt}. We now consider the function $H$ with $\tau(u) = 0.$ The classes $A, B, C$ and $D$ become respectively ${\it Aiv, Biv, Civ}$ and ${\it Div}$. The NGSs, gauge functions and first integrals of these cases are given in Table \ref{T1}.

\begin{table}[!ht]
  \centering
  \caption{ NGSs, gauge functions and first integrals for the classes $A$, $B$, $C$, $D$, $Aiv$, $Civ$ and $Div$. The vector fields ${\bf Z}_1  = 2 m v \partial_v + m r \partial_r + 2\partial_{\theta}$ and ${\bf Z}_2 = 2 u \partial_u + r \partial_r$ in classes $A, Aiv, A(c), C(a), Civ$ and $Div$ is a \emph{HKV}. \label{T1} }
  \resizebox{17cm}{!} {
\begin{tabular}{llll}
\hline\hline
 \,\,\, Class & \qquad NGS & $ f $ & First Integrals \\
 (subcase) &  &  &  \\
\hline

A & ${\bf k} = \partial_v$, \quad ${\bf Y}_1 = \partial_s$, \quad ${\bf  Y}_2 = - s \partial_v$ & $0$, \quad $0$, \quad $u$ & $I_{1} = - \dot{u}$, \quad $I_{2} = - E_{L}$, \quad $I_{3} = s \dot{u} - u$ \\

  & ${\bf Y}_3 = 2 m s \partial_s + {\bf Z}_1$ & $0$ & $I_{4} = - 2 m s E_{L} - 2 m v \dot{u} + m r \dot{r} + 2 r^2 {\dot{\theta}^2}$, \quad  \\
   & & & \\
  A(a) & ${\bf k} = \partial_v$, \quad ${\bf Y}_1 = \partial_s$, \quad ${\bf  Y}_2 = - s \partial_v$ & $0$, \quad $0$, \quad $u$ & $I_{1} = - \dot{u}$, \quad $I_{2} = - E_{L}$, \quad $I_{3} = s \dot{u} - u$ \\
   & & & \\
 A(b) & ${\bf k} = \partial_v$, \quad ${\bf Y}_1 = \partial_s$, \quad ${\bf  Y}_2 = - s \partial_v$ & $0$, \quad $0$, \quad $u$ & $I_{1} = - \dot{u}$, \quad $I_{2} = - E_{L}$, \quad $I_{3} = s \dot{u} - u$ \\

      & ${\bf X}_2 = \partial_u$ & $0$ & $I_{4} = -2 H \dot{u} - \dot{v}$ \\
   & & & \\
  A(c) & ${\bf k} = \partial_v$, \quad ${\bf Y}_1 = \partial_s$, \quad ${\bf  Y}_2 = - s \partial_v$ & $0$, \quad $0$, \quad $u$ & $I_{1} = - \dot{u}$, \quad $I_{2} = - E_{L}$, \quad $I_{3} = s \dot{u} - u$ \\

       & ${\bf Y}_3 = 2 s \partial_s + {\bf Z}_2$ & $0$ & $I_{4} = -2 s E_{L} - 4 H u \dot{u} - 2 u \dot{v} + r \dot{r}$ \\
\hline

 B(a) & ${\bf k} = \partial_v$, \quad ${\bf Y}_1 = \partial_s$, \quad ${\bf  Y}_2 = - s \partial_v$ & $0$, \quad $0$, \quad $u$ & $I_{1} = - \dot{u}$, \quad $I_{2} = - E_{L}$, \quad $I_{3} = s \dot{u} - u$ \\

      & ${\bf X}_2 = G'_1 P_1(\theta) r \partial_v + G_1 P_1(\theta) \partial_r - \frac{G_1}{r} P_2(\theta) \partial_{\theta}$ & $0$ & $I_{4} = - G'_{1} P_1(\theta) r \dot{u} + G_{1} P_1(\theta) \dot{r} - G_{1} P_2(\theta) r \dot{\theta}$ \\

      & ${\bf X}_3 = G'_2 P_1(\theta) r \partial_v + G_2 P_1(\theta) \partial_r - \frac{G_2}{r} P_2(\theta) \partial_{\theta}$ & $0$& $I_{5} = - G'_{2} P_1(\theta) r \dot{u} + G_{2}P_1(\theta) \dot{r} - G_{2} P_2(\theta) r \dot{\theta}$ \\
   & & & \\
  B(b) & ${\bf k} = \partial_v$, \quad ${\bf Y}_1 = \partial_s$, \quad ${\bf  Y}_2 = - s \partial_v$ & $0$, \quad $0$, \quad $u$ & $I_{1} = - \dot{u}$, \quad $I_{2} = - E_{L}$, \quad $I_{3} = s \dot{u} - u$ \\

       & ${\bf X}_2 = \partial_u$ & $0$ & $I_{4} = - 2 H \dot{u} - \dot{v}$ \\

       & ${\bf X}_3 = D'_1 P_1(\theta) r \partial_v + D_1 P_1(\theta) \partial_r - \frac{D_1}{r} P_2(\theta) \partial_\theta$ & $0$ & $I_{5} = - D'_{1} P_1(\theta) r \dot{u} + D_{1} P_1(\theta) \dot{r} - D_{1} P_2(\theta) r \dot{\theta}$ \\

       & ${\bf X}_4 = D'_2 P_1(\theta) r \partial_v + D_2 P_1(\theta) \partial_r - \frac{D_2}{r} P_2(\theta) \partial_\theta$ & $0$ & $I_{6} = - D'_{2} P_1(\theta) r \dot{u} + D_{2} P_1(\theta) \dot{r} - D_{2} P_2(\theta) r \dot{\theta}$ \\

\hline

C & ${\bf k} = \partial_v$, \quad ${\bf Y}_1 = \partial_s$, \quad ${\bf  Y}_2 = - s \partial_v$ & $0$, \quad $0$, \quad $u$ & $I_{1} = - \dot{u}$, \quad $I_{2} = - E_{L}$, \quad $I_{3} = s \dot{u} - u$ \\

  & ${\bf X}_2 = \partial_\theta$ & $0$ & $I_{4} = r^2 \dot{\theta}$ \\
   & & & \\
C(a) & ${\bf k} = \partial_v$, \quad ${\bf Y}_1 = \partial_s$, \quad ${\bf  Y}_2 = - s \partial_v$ & $0$, \quad $0$, \quad $u$ & $I_{1} = - \dot{u}$, \quad $I_{2} = - E_{L}$, \quad $I_{3} = s \dot{u} - u$ \\

     & ${\bf X}_2 = \partial_\theta$, \quad ${\bf Y}_3 = 2 s \partial_s + {\bf Z}_2$ & $0$, \quad $0$ & $I_{4} = r^2 \dot{\theta}$, \quad $I_{5} = -2 s E_{L} - 4 H u \dot{u} - 2 u \dot{v} + r \dot{r}$ \\

\hline

 D & ${\bf k} = \partial_v$, \quad ${\bf Y}_1 = \partial_s$, \quad ${\bf  Y}_2 = - s \partial_v$ & $0$, \quad $0$, \quad $u$ & $I_{1} = - \dot{u}$, \quad $I_{2} = - E_{L}$, \quad $I_{3} = s \dot{u} - u$ \\

\hline

 Aiv & ${\bf k} = \partial_v$, \quad ${\bf Y}_1 = \partial_s$, \quad ${\bf  Y}_2 = - s \partial_v$ & $0$, \quad $0$, \quad $u$ & $I_{1} = - \dot{u}$, \quad $I_{2} = - E_{L}$, \quad $I_{3} = s \dot{u} - u$ \\

     & ${\bf X}_2 = \partial_u$, & $0$, & $I_{4} = - 2 H \dot{u} - \dot{v}$, \\
     & ${\bf X}_3 = m (u \partial_u - v \partial_v) - \partial_\theta$ & $0$ & $I_{5} =  - m u (2 H \dot{u} + \dot{v}) + m v \dot{u} - r^2 \dot{\theta}$ \\
     & ${\bf Y}_3 = 2 m s \partial_s + {\bf Z}_1$ & $0$ & $I_{6} = -2 m s E_{L} - 2 m v \dot{u} + m r \dot{r} + 2 r^2 \dot{\theta}$ \\

\hline

 Civ & ${\bf k} = \partial_v$, \quad ${\bf Y}_1 = \partial_s$, \quad ${\bf  Y}_2 = - s \partial_v$ & $0$, \quad $0$, \quad $u$ & $I_{1} = - \dot{u}$, \quad $I_{2} = - E_{L}$, \quad $I_{3} = s \dot{u} - u$ \\

     & ${\bf X}_2 = \partial_u$, \quad ${\bf X}_3 = \partial_\theta$ & $0$, \quad $0$ & $I_{4} = -2 H \dot{u} - \dot{v}$, \quad $I_{5} = r^2 \dot{\theta}$ \\

     & ${\bf Y}_3 = 2 s \partial_s + {\bf Z}_2$ & $0$ & $I_{6} = -2 s E_{L} - 4 H u \dot{u} - 2 u \dot{v} + r \dot{r}$ \\

\hline

Div & ${\bf k} = \partial_v$, \quad ${\bf Y}_1 = \partial_s$, \quad ${\bf  Y}_2 = - s \partial_v$ & $0$, \quad $0$, \quad $u$ & $I_{1} = - \dot{u}$, \quad $I_{2} = - E_{L}$, \quad $I_{3} = s \dot{u} - u$ \\

    & ${\bf X}_2 = \partial_u$, & $0$, & $I_{4} = - 2 H \dot{u} - \dot{v}$, \\
    & ${\bf Y}_3 = 2 s \partial_s + {\bf Z}_2$ & $0$ & $I_{5} = -2 s E_{L} - 4 H u \dot{u} - 2 u \dot{v} + r \dot{r}$ \\
\hline\hline
\end{tabular}
}
\end{table}

\begin{table}[!ht]
  \centering
\caption{ Noether gauge symmetry algebras for pp-wave spacetimes. \label{T2}}
  \resizebox{17cm}{!} {
\begin{tabular}{llll}
\hline\hline
 \,\,\, Class & Algebra & Metric Function $ H $ & $ F $ \\
 (subcase) &  &  &  \\
\hline
  A & $\mathcal{N}_4 \supset \mathcal{H}_2 \supset \mathcal{G}_1$ & $\tau(u) r^2 + l e^{2 m \theta} r^{-2}$ & $4 \tau(u) + 4 l (1 + m^2) e^{2 m \theta} r^{-4}$ \\
  A(a) & $\mathcal{N}_4 \supset \mathcal{G}_2$ & $w r^2 + l e^{2 m \theta} r^{-2}$ & $4w + 4 l (1 + m^2) e^{2 m \theta} r^{-4}$ \\
  A(b) & $\mathcal{N}_4 \supset \mathcal{H}_2 \supset \mathcal{G}_1$ & $w u^{-2} r^2 + l e^{2 m \theta} r^{-2}$ & $4 w u^{-2} + 4 l (1 + m^2) e^{2 m \theta} r^{-4}$ \\
  B(a) & $\mathcal{N}_5 \supset \mathcal{G}_3$ & $\frac{q^2}{8} u^{-4} r^2 + l (\sigma z - \rho y)^{-2}$ & $ \frac{q^2}{2} u^{-4} + 6 l (\sigma^2 + \rho^2) (\sigma z - \rho y)^{-4}$ \\
  B(b) & $\mathcal{N}_6 \supset \mathcal{G}_4$ & $q r^2 + l (\sigma z - \rho y)^{-2}$ & $4 q + 6 l (\sigma^2 + \rho^2) (\sigma z - \rho y)^{-4}$ \\
  C & $\mathcal{N}_4 \supset \mathcal{G}_2$ & $\tau(u) r^2 + \delta r^{-2}$ & $4 \tau(u) + 4 \delta r^{-4}$ \\
  C(a) & $\mathcal{N}_5 \supset \mathcal{H}_3 \supset \mathcal{G}_2$ & $w u^{-2} r^2 + \delta r^{-2}$ & $4 w u^{-2} + 4 \delta r^{-4}$ \\
  D & $\mathcal{N}_3 \supset \mathcal{G}_1$ & $\tau(u) r^2 + \delta(\theta) r^{-2}$ & $4 \tau(u) + (4 \delta(\theta) + \delta(\theta)_{,\theta \theta}) r^{-4}$ \\
  Aiv & $\mathcal{N}_6 \supset \mathcal{H}_4 \supset \mathcal{G}_3$ & $l e^{2 m \theta} r^{-2}$ & $4 l e^{2 m \theta} (1 + m^2) r^{-4}$ \\
  Biv & $\mathcal{N}_8 \supset \mathcal{H}_5 \supset \mathcal{G}_4$ & $l (\sigma z - \rho y)^{-2} $ & $6 l (\sigma^2 + \rho^2) (\sigma z - \rho y)^{-4}$ \\
  Civ & $\mathcal{N}_6 \supset \mathcal{H}_4 \supset \mathcal{G}_3$ & $\delta r^{-2}$ & $4 \delta r^{-4}$ \\
  Div & $\mathcal{N}_5 \supset \mathcal{H}_3 \supset \mathcal{G}_2$ & $\delta(\theta) r^{-2}$ & $(4 \delta(\theta) + \delta(\theta)_{,\theta \theta}) r^{-4}$ \\
\hline\hline
\end{tabular}
}
\end{table}

\noindent In this subsection, we have mostly give the obtained NGSs, gauge functions and first integrals in Table \ref{T8}. Now we consider the isometry classes of Sippel and Goenner \cite{sg} in order to obtain any further NGSs if they admit them (see also Table \ref{T3}).

\begin{table}[!ht]
  \centering
\caption{Noether gauge symmetries for Sippel and Goenner [6] isometry classes.\label{T3}}
  \resizebox{17cm}{!} {
\begin{tabular}{llll}
\hline\hline
 Class & Algebra & Metric Function $ H $ & $ F $ \\
\hline
  1 & $\mathcal{N}_3 \supset \mathcal{G}_1$ & $Arbitrary$ & $H_{,yy} + H_{,zz}$ \\
  1i & $\mathcal{N}_6 \supset \mathcal{G}_3$ & $H(u,z)$ & $H_{,zz}$ \\
  2 & $\mathcal{N}_4 \supset \mathcal{G}_2$ & $H(u,r)$ & $H_{,rr} + r^{-1} H_{,r}$ \\
  2i & $\mathcal{N}_5 \supset \mathcal{H}_3 \supset \mathcal{G}_2$ & $K e^{\alpha u} \ln|r|$ & 0 \\
  2ii & $\mathcal{N}_5 \supset \mathcal{H}_3 \supset \mathcal{G}_2$ & $K (\alpha u + \beta)^q$ & 0 \\
  2iii & $\mathcal{N}_6 \supset \mathcal{H}_4 \supset \mathcal{G}_3$ & $K \ln|r|$ & 0 \\
  3 & $\mathcal{N}_4 \supset \mathcal{G}_2$ & $u^{-2}W(s,t)$ & $u^{-2}(W_{,ss} + W_{,tt})$ \\
  4 & $\mathcal{N}_4 \supset \mathcal{G}_2$ & $W(s,t)$ & $W_{,ss} + W_{,tt}$ \\
  5 & $\mathcal{N}_5 \supset \mathcal{G}_3$ & $u^{-2} W(r)$ & $u^{-2} (W_{,rr} + r^{-1} W_{,r})$ \\
  5i & $\mathcal{N}_5 \supset \mathcal{G}_3$ & $u^{-2} \zeta \ln|r|$ & 0 \\
  5ii & $\mathcal{N}_5 \supset \mathcal{G}_3$ & $u^{-2} \delta r^{-\sigma} $ & $u^{-2} (\delta \sigma^2 r^{-(\sigma + 2)}-4 \sigma (2 - \sigma)^{-2})$ \\
 & & $\,\,\, - u^{-2} \sigma (2 - \sigma)^{-2} r^2$ & \\
  6 & $\mathcal{N}_5 \supset \mathcal{G}_3$ & $W(r)$ & $W_{,rr} + r^{-1} W_{,r}$ \\
  6i & $\mathcal{N}_6 \supset \mathcal{H}_4 \supset \mathcal{G}_3$ & $\delta r^{-\sigma}$ & $\delta \sigma^2 r^{-(\sigma + 2)}$ \\
  6ii & $\mathcal{N}_5 \supset \mathcal{G}_3$ & $\zeta \ln|r|$ & 0 \\
  7 & $\mathcal{N}_5 \supset \mathcal{G}_3$ & $e^{2 c \theta} W(r)$ & $e^{2 c \theta} (W_{,rr} + r^{-1} W_{,r} + 4 c^2 r^{-2} W)$ \\
  7i & $\mathcal{N}_6 \supset \mathcal{H}_4 \supset \mathcal{G}_3$ & $\delta r^{-\sigma} e^{2 c \theta}$ & $\delta e^{2 c \theta} (\sigma^2 + 4 c^2) r^{-(\sigma + 2)}$ \\
  8 & $\mathcal{N}_7 \supset \mathcal{G}_3$ & $W(t)e^{2\delta w}$ & $[(\eta^2 + \sigma^2)W_{,ss} + 4\epsilon^2(\eta^2 + \sigma^2)^{-1}W]e^{2t}$ \\
  8($\epsilon = 0$) & $\mathcal{N}_7 \supset \mathcal{G}_4$ & $W(t)$ & $[(\eta^2 + \sigma^2)W_{,ss} + 4\epsilon^2(\eta^2 + \sigma^2)^{-1}W]e^{2t}$ \\
  8($\epsilon = 0$)i & $\mathcal{N}_8 \supset \mathcal{H}_5 \supset \mathcal{G}_4$ & $K t^c, c \neq -2$ & $[(\eta^2 + \sigma^2)W_{,ss} + 4\epsilon^2(\eta^2 + \sigma^2)^{-1}W]e^{2t}$ \\
  9 & $\mathcal{N}_8 \supset \mathcal{G}_5$ & $K e^{2 (\eta y - \sigma z)}$ & $4 (\eta^2 + \sigma^2) H$ \\

\hline\hline
\end{tabular}
}
\end{table}

\begin{table}[!ht]
  \centering
  \caption{ NGSs, gauge functions and first integrals for the classes $2$, $5$, $6$ and $7$. \label{T8} }
  \resizebox{17cm}{!} {
\begin{tabular}{llll}
\hline\hline
 \,\,\, Class & \qquad NGS & $ f $ & First Integrals \\
 (subcase) &  &  &  \\

\hline

2 & ${\bf k} = \partial_v$, \quad ${\bf Y}_1 = \partial_s$, \quad ${\bf  Y}_2 = - s \partial_v$ & $0$, \quad $0$, \quad $u$ & $I_{1} = - \dot{u}$, \quad $I_{2} = - E_{L}$, \quad $I_{3} = s \dot{u} - u$ \\

  & ${\bf X}_2 = \p_\theta$ & $0$ & $I_{4} = r^2 \dot{\theta}$ \\

\hline

5 & ${\bf k} = \partial_v$, \quad ${\bf Y}_1 = \partial_s$, \quad ${\bf  Y}_2 = - s \partial_v$ & $0$, \quad $0$, \quad $u$ & $I_{1} = - \dot{u}$, \quad $I_{2} = - E_{L}$, \quad $I_{3} = s \dot{u} - u$ \\

  &${\bf X}_2 = u \partial_u - v \partial_v$, \quad ${\bf X}_3 = \partial_\theta$ & $0$, \quad $0$ & $I_{4} = - (2 H \dot{u} + \dot{v}) + v \dot{u}$, \quad $I_{5} = r^2 \dot{\theta}$ \\

\hline

6 & ${\bf k} = \partial_v$, \quad ${\bf Y}_1 = \partial_s$, \quad ${\bf  Y}_2 = - s \partial_v$ & $0$, \quad $0$, \quad $u$ & $I_{1} = - \dot{u}$, \quad $I_{2} = - E_{L}$, \quad $I_{3} = s \dot{u} - u$ \\

  & ${\bf X}_2 = \partial_u$, \quad ${\bf X}_3 = \partial_\theta$ & $0$, \quad $0$ & $I_{4} = - 2 H \dot{u} - \dot{v}$, \quad $I_{5} = r^2 \dot{\theta}$ \\

\hline

7 & ${\bf k} = \partial_v$, \quad ${\bf Y}_1 = \partial_s$, \quad ${\bf  Y}_2 = - s \partial_v$ & $0$, \quad $0$, \quad $u$ & $I_{1} = - \dot{u}$, \quad $I_{2} = - E_{L}$, \quad $I_{3} = s \dot{u} - u$ \\

  & ${\bf X}_2 = \partial_u$, \quad ${\bf X}_3 = c (u \partial_u - v \partial_v) - \partial_\theta$ & $0$, \quad $0$ & $I_{4} = - 2 H \dot{u} - \dot{v}$, \quad $I_{5} = - c u ( 2 H \dot{u} + \dot{v}) + c v \dot{u} - r^2 \dot{\theta}$  \\

\hline\hline
\end{tabular}
}
\end{table}

\subsubsection{Isometry class 2.}

In this class, Sippel and Goenner \cite{sg} give a differential equation for the function $H$ as $z H_{,y} - y H_{,z} = 0$ which yields the form $H_{,\theta} = 0$ for polar coordinates. Therefore the function $H$ depends only on the variables $u$ and $r$, i.e. $H = H(u,r)$. The conditions (\ref{ceq1p}) and (\ref{ceq2p}) in polar coordinates give $c_3 = 0, c_4 = 0$ and $g_1 = a_1 u + a_2$, where $a_1$ and $a_2$ are constants, and
\begin{eqnarray}
& & r g_{2,uu} + H_{,r} g_2 = 0, \label{ic2-1}\\& & r g_{3,uu} + H_{,r} g_3 = 0, \label{ic2-2}\\& & (c_6 u + c_7) H_{,u} + c_1 \frac{r}{2} H_{,r} + (2 c_6 -c_1) H + g_{4,u} = 0. \label{ic2-3}
\end{eqnarray}
For an arbitrary $H(u,r)$ we get from the above equations that the NGSs are the KVs already listed in \cite{sg}, i.e. ${\bf k}, {\bf X}_2, {\bf Y}_1$, and ${\bf Y}_2$, which means that the spacetime admits $\mathcal{N}_4 \supset \mathcal{G}_2$.

Let us assume that $H= m(u) r^p + n(u) r^q$, where $p\neq q$ are constants. In order to avoid plane wave spacetimes, one of $p,q$ must equal to $2$. Therefore we choose $q=2$ and assume $p \neq 0,2$ (otherwise type O). Substituting the function $H= m(u) r^p + n(u) r^2$, ($p \neq 0,2$), into Eqs. (\ref{ic2-1}), (\ref{ic2-2}) and (\ref{ic2-3}), it can be found that $m(u) = \sigma u^{\alpha}, n(u) = \eta u^{-2}, c_6 = c_1 (2-p)/ (2 \alpha + 4), g_4 (u) = a_3$ and $g_2 (u) =g_3 (u) = c_7 = 0$ which means
\begin{eqnarray}
& &  \xi = c_1 s + c_2, \,\, \eta^1 = c_1 \frac{(2-p)}{(2 \alpha + 4)} u, \,\, \eta^2 = -a_1 s + c_1 \frac{(p+ 2 \alpha +2)}{(2 \alpha + 4)} v + a_3, \\& & \eta^3 = c_1 \frac{r}{2}, \,\, \eta^4 = c_5, \,\, f = a_1 u + a_2, \nonumber
\end{eqnarray}
where $\alpha, \sigma, \eta, a_3$ are constants. Hence, we have five NGS, that is, in addition to the two KVs ${\bf k}, {\bf X}_2$ and two non-special NGSs ${\bf Y}_1, {\bf Y}_2$ given above, there is a special NGS as ${\bf Y}_3 = 2 s \partial_s + {\bf Z}, \label{nons-ngs}$ with the first integral $I_{5} = 2 I_2 s + \frac{I_1}{\alpha +2} \left[ (2-p) H u + (p+ 2 \alpha + 2) v \right] + \frac{(p-2)}{2(\alpha + 2)} u \dot{v} + r \dot{r}$, where ${\bf Z}$ is the proper HKV of the form
\begin{equation}
{\bf Z} = \frac{(2-p)}{(\alpha +2)} u \partial_u + \frac{(p+ 2 \alpha +2)}{(\alpha +2)} v \partial_v + r \partial_r, \quad  q\neq 0,2, \,\, \alpha \neq -2.
\end{equation}
Here the algebra is $\mathcal{N}_5 \supset \mathcal{H}_3 \supset \mathcal{G}_2$.

Now let us consider the vacuum specialization cases that includes the metric function of the form $H = e^{g(u)} \ln|r|$. The additional KV can appear only in the cases $H= K\, \ln |r|$ and $H = K (\alpha u + \beta )^{-2} \ln |r|$, and in both cases they admit $\mathcal{N}_6 \supset \mathcal{H}_4 \supset \mathcal{G}_3$ (isometry class 6 specialization) and $\mathcal{N}_5 \supset \mathcal{G}_3$ (isometry class 5 specialization) algebras, respectively. The additional KV to ${\bf k}$ and ${\bf X}_2$ is ${\bf X}_3 = \partial_u$ for $H= K\, \ln |r|$, and ${\bf X}_3 = (\alpha u + \beta) \partial_u- \alpha v \partial_v$ for $H = K (\alpha u + \beta )^{-2} \ln |r|$.

There are three possibilities in which the NGSs include proper HKV \cite{kt}: \\
(i) $\mathcal{N}_5 \supset \mathcal{H}_3 \supset \mathcal{G}_2$, i.e. $H= K e^{\alpha u} \ln|r|, \, \alpha \neq 0$, with non-special NGS of the form as (\ref{nons-ngs}) given above, where ${\bf Z}$ is the proper HKV as follows
\begin{equation}
{\bf Z} = \frac{2}{\alpha } \partial_u + (2 v - \frac{K}{\alpha} e^{\alpha u} ) \partial_v + r \partial_r.
\end{equation}
(ii) $\mathcal{N}_5 \supset \mathcal{H}_3 \supset \mathcal{G}_2$,  $\alpha \neq 0$, $\beta$ arbitrary, $q \neq -2$,  $H= K (\alpha u + \beta)^q \ln|r|$. For the case with $q \neq -1,-2$ the non-special NGS is same as (\ref{nons-ngs}) but proper HKV has the form
\begin{equation}
{\bf Z} = \frac{2}{\alpha (q + 2)} (\alpha u + \beta) \partial_u + \left[ 2 \frac{(q+1)}{(q+2)} v - \frac{K (\alpha u + \beta)^{q+1}}{\alpha (q+ 1)} \right] \partial_v + r \partial_r,
\end{equation}
and for the case with $q = -1$ the NGS has the proper HKV
\begin{equation}
{\bf Z} = \frac{2}{\alpha } (\alpha u + \beta) \partial_u  - \frac{K }{\alpha} \ln (\alpha u + \beta) \partial_v + r \partial_r.
\end{equation}
(iii) $\mathcal{N}_6 \supset \mathcal{H}_4 \supset \mathcal{G}_3$,  i.e. $H= K\, \ln|r|$. In this case, the special NGS ${\bf Y}_3$ has the form of (\ref{nons-ngs}) and includes proper HKV
\begin{equation}
{\bf Z} = u \partial_u  + (v - K u ) \partial_v + r \partial_r.
\end{equation}

\subsubsection{Isometry class 5.}

In this case, Sippel and Goenner \cite{sg} give differential equations for $H$ as follows
\begin{equation}
u H_{,u} + 2 H = 0, \qquad z H_{,y} - y H_{,z} =0. \label{ic5-deq}
\end{equation}
The latter one of these differential equations yields $H_{,\theta} =0$, that is $H=H(u,r)$ in polar coordinates. Thus using the first one of Eq.(\ref{ic5-deq}) the function $H$ has the form $H = u^{-2} W(r)$. For an arbitrary form of $W(r)$ in which we exclude the type \emph{O} pp-wave spacetimes, the NGSs are ${\bf k}, {\bf X}_2, {\bf X}_3, {\bf Y}_1$ and ${\bf Y}_2$.

In this class there are two subcases: {\it (i)} $H = u^{-2} \zeta \ln|r|$ and {\it (ii)} $H = u^{-2} (\delta r^{-\sigma} - \sigma (2 - \sigma)^{-2} r^2)$. The NGSs, Lie brackets and the first integrals of those subcases are same with the Class 5.

\subsubsection{Isometry class 6.}

The constraint differential equations of \cite{sg} for $H$ in this class are given by
\begin{equation}
H_{,u} = 0, \qquad z H_{,y} - y H_{,z} =0. \label{ic6-deq}
\end{equation}
The second one gives rise to $H_{,\theta} =0$, and so the function $H$ has the form $H = W(r)$.
For this case, the constraint equations (\ref{ceq1p}) and (\ref{ceq2p}) become
\begin{eqnarray}
& & \qquad g_{1,uu} + \left( c_3 \sin \theta - c_4 \cos \theta \right) W_{,r} = 0, \label{ceq1p-ic6} \\& & \qquad \cos \theta \left( r g_{2,uu} + W_{,r} g_2 \right) + \sin \theta \left( r g_{3,uu} + W_{,r} g_{3} \right) + g_{4,u}  \nonumber \\& & \qquad \qquad \qquad \qquad + \frac{c_1}{2} \left( r W_{,r} - 2 W \right) +  2 c_6 W = 0. \label{ceq2p-ic6}
\end{eqnarray}
If $W(r)$ is an arbitrary function of $r$ then it follows immediately from (\ref{ceq1p-ic6}) and (\ref{ceq2p-ic6}) that the NGSs are the three KV ${\bf k}, {\bf X}_2, {\bf X}_3$ and two non-special NGS ${\bf Y}_1, {\bf Y}_2$, Thus the Noether algebra is at least $\mathcal{N}_5 \supset \mathcal{G}_3$.

When $W_{,r} \neq constant \neq 0$, then the constraint differential equation (\ref{ceq1p-ic6}) leads to $c_2 =c_3 =0$ and $g_{1,uu}=0$ which give $g_1 = a_1 u + a_2$, where $a_1,a_2$ are constants, and  the constraint differential equation (\ref{ceq2p-ic6}) yields
\begin{eqnarray}
& & r g_{2,uu} + W_{,r} g_2 = 0, \label{ceq1-ic6}\\& &  r g_{3,uu} + W_{,r} g_3 = 0, \label{ceq2-ic6} \\& & g_{4,u} + c_1 \frac{r}{2} W_{,r} + (2 c_6 -c_1 ) W  = 0. \label{ceq3-ic6}
\end{eqnarray}
If $W_{,r}/r = constant$, then we get plane wave spacetimes which will be considered in the following section. If $H$ has the form $H=W(r)=\delta r^{-\sigma}$, we find from the above constraint equations that the spacetime admits an $\mathcal{N}_6 \supset \mathcal{H}_4 \supset \mathcal{G}_3$ consisting of ${\bf k}, {\bf X}_2,{\bf X}_3,{\bf Y}_1,{\bf Y}_2$ and the special NGS ${\bf Y}_3 = 2 s \p_s + {\bf Z}$ where ${\bf Z}$ is the proper HKV which have the form ${\bf Z} = \frac{(2+\sigma)}{2} u \p_u + \frac{(2-\sigma)}{2} v \p_v + r \p_r$. The corresponding first integral for ${\bf Y}_3$ is $I_{6} = -2 s E_{L} + \frac{(2 - \sigma)}{2} u (2 H \dot{u} + \dot{v}) - \frac{(2 + \sigma)}{2} v \dot{u} + r \dot{r}$.

If $H = \zeta \ln|r|$, which form of $H$ exclude type \emph{O}, then Eq.(\ref{ceq1p-ic6}) gives $c_2$ and $c_3$ vanish, and $g_1 = a_1 u + a_2$. For this $H$, Eqs. (\ref{ceq1-ic6}) and (\ref{ceq2-ic6}) imply $g_2=g_3 =0$. The Eq.(\ref{ceq3-ic6}) integrates to give $c_6 =c_1 /2$ and $g_4 = -c_1 \zeta /2 + a_3$. This spacetime admits also $\mathcal{N}_6 \supset \mathcal{H}_4 \supset \mathcal{G}_3$ consisting of three KVs ${\bf k}, {\bf X}_2, {\bf X}_3$, two non-special NGSs ${\bf Y}_1, {\bf Y}_2$ and one special NGS ${\bf Y}_3 = 2 s \partial_s + {\bf Z}$ where ${\bf Z}$ is the proper HKV ${\bf Z} = u \partial_u + ( v- \zeta u) \partial_v + r \partial_r$.

\subsubsection{Isometry class 7.}

The differential equations of this case given by Sippel and Goenner \cite{sg} in polar coordinates are
\begin{equation}
H_{,u} = 0, \qquad  H_{,\theta} - 2 c H =0. \label{ic7-deq}
\end{equation}
which yields $H = e^{2 c \theta} W(r)$ (see also Ref.\cite{kt}). Comparing these differential equations with constraint equation (\ref{ceq2p}), we observe that $c_5 = k_1$ and $c_6 = -c k_1$, where $k_1$ is an arbitrary constant parameter.

If $W$ is an arbitrary function then one finds immediately that the only NGS which occur are the three KV given in \cite{sg} and two non-special NGS, i.e., ${\bf k}, {\bf X}_2, {\bf X}_3, {\bf Y}_1$ and ${\bf Y}_2$.

If W(r) is not arbitrary function of $r$, then the constraint equation (\ref{ceq2p}) gives $W(r)=\delta r^{-2}$, where $\delta$ is a non-zero constant. Then the constraint equations (\ref{ceq1p}) and (\ref{ceq2p}) lead to an $\mathcal{N}_6 \supset \mathcal{H}_4 \supset \mathcal{G}_3$ with basis ${\bf k}, {\bf X}_2, {\bf X}_3, {\bf Y}_1, {\bf Y}_2$ and the special NGS ${\bf Y}_3 = 2 s \p_s + {\bf Z}$ with the first integral $I_{6} = -2 s E_{L} - 2 v \dot{u} + r \dot{r}$ where ${\bf Z}$ is the proper HKV and ${\bf Z} = 2 v \p_v + r \p_r$.

\section{Noether gauge symmetries for plane wave spacetimes}
\label{planewave}

When the function $H$ has the form given by (\ref{plane-wave}), then pp-wave spacetime (\ref{pp-wave}) represents an Einstein-Maxwell plane wave spacetime and admits at least an $\mathcal{H}_6$. When the plane wave spacetime is vacuum (i.e. $A(u)=-C(u)$) or conformally flat (i.e. $A(u) = C(u)$ and $B(u)=0$), it admits an $\mathcal{H}_7$ subalgebra. Now we wish to find the maximum number of NGS admitted by the plane wave spacetimes.

For the plane wave spacetime classes in \cite{kt}, we calculated the gauge function $f$, and the components $\xi, \eta^1, \eta^2, \eta^3$ and $\eta^4$ of NGS vector field in Eqs.(\ref{xieta1})-(\ref{f}) with the constraint equations (\ref{ceq1}) and (\ref{ceq2}) which lead to the following equations
\begin{eqnarray}
& & f_{1,uu} =0, \quad f_{4,u} =0, \label{ceq0-pw}\\& & f_{2,uu} + A(u) f_2 + B(u) f_3 = 0, \label{ceq1-pw} \\& & f_{3,uu} + C(u) f_3 + B(u) f_2 = 0, \label{ceq2-pw}  \\ & &  c_3 A(u) + c_4 B(u) =0, \quad c_3 B(u) + c_4 C(u) =0, \label{ceq3-pw} \\& & (c_6 u + c_7) A'(u) + 2 c_6 A(u) - 2 c_5 B(u) =0, \label{ceq4-pw} \\& & (c_6 u + c_7) B'(u) + 2 c_6 B(u) + c_5 \left[ A(u) - C(u) \right] =0, \label{ceq5-pw} \\& & (c_6 u + c_7) C'(u) + 2 c_6 C(u) + 2 c_5 B(u) =0, \label{ceq6-pw}
\end{eqnarray}
where $c_i$ ($i=3,...,7$) are five constant parameters, and $f_1, f_2, f_3, f_4$ are functions of $u$. As it was stated in section \ref{sec2-ngs}, the constant parameter $c_2$, the functions $f_1(u)$ and $f_4 (u)$ give rise to the KV ${\bf k}$ and two additional non-special NGS ${\bf Y}_1$ and ${\bf Y}_2$ given in (\ref{vf12}) and (\ref{vf3}),  and so we shall put $c_2, f_1$ and $f_4$ to zero. The equations (\ref{ceq1-pw}) and (\ref{ceq2-pw}) correspond to the condition on $H$ to admit the \emph{four} KV listed in case 10 of \cite{sg}
\begin{eqnarray}
& & {\bf X}_i = f_2 (u)_i \p_y + f_3 (u)_i \p_z + \left[ y (f_{2,u})_i + z (f_{3,u})_i \right] \p_v, \qquad (i=2,...,5)
\end{eqnarray}
and so we can put $f_2 (u)=0$ and $f_3(u)=0$ for the remaining NGS. Furthermore, since the $c_1$ parameter is not appear in the constraint equations (\ref{ceq0-pw})-(\ref{ceq6-pw}) it gives rise to a non-special NGS vector field for any form of $H$ in plane wave spacetimes as
\begin{equation}
{\bf Y}_3 = 2 s \p_s + {\bf Z},
\end{equation}
where ${\bf Z}= 2 v \p_v + y\p_y + z\p_z$ is a proper HKV. Thus, all plane wave spacetimes admit at least an $\mathcal{N}_8 \supset \mathcal{H}_6 \supset \mathcal{G}_5$. Now we need to consider the following NGS vector field for non-conformally flat or conformally flat spacetimes:
\begin{eqnarray}
& & {\bf Y} = s ( c_3 \p_y + c_4 \p_z )+ c_5 (z \p_y -y \p_z) + c_6 ( u \p_u - v \p_v ) + c_7 \p_u,
\end{eqnarray}
with gauge function $f=c_3 y +c_4 z$.

\subsection{Non-conformally flat spacetimes}

Assuming $A(u) \neq C(u)$ and $B(u) \neq 0$ for non-conformally flat spacetimes there are some specializations of those functions $A(u), B(u), C(u)$ in Eqs. (\ref{ceq3-pw})-(\ref{ceq6-pw}) which yields one additional NGS. We will consider such specializations, four of these due to the cases 11-14 of table 2 in \cite{sg} which give rise to an additional KV, and two case of \cite{tupper2003}.

(i) {\it Isometry class 11} of \cite{sg}. The functions $A(u), B(u), C(u)$ of this case have the form
\begin{eqnarray}
& & A(u) = a u^{-2}, \quad B(u) = b u^{-2}, \quad C(u) = c u^{-2},
\end{eqnarray}
where $a, b$ and $c$ are constants. In addition to five KVs ${\bf k}, X_i$, and three NGSs ${\bf Y}_1, {\bf Y}_2, {\bf Y}_3$, in which ${\bf Y}_3$ includes the HKV ${\bf Z}$, there exist one additional KV
\begin{eqnarray}
& & {\bf X}_6 = u \p_u - v \p_v,
\end{eqnarray}
and one special NGS
\begin{eqnarray}
& & {\bf Y}_4 = s (\sigma \p_y + \eta \p_z ),
\end{eqnarray}
where $c_3 \equiv \sigma, c_4 \equiv \eta, b = - (\sigma / \eta) a$ and $c= (\sigma^2 / \eta^2) a$. Hence in this case the spacetime admits $\mathcal{N}_{10} \supset \mathcal{H}_7 \supset \mathcal{G}_6$.

(ii) {\it Isometry class 12} of \cite{sg}. For this case
\begin{equation}
A(u) = c u^{-2} \left( \sin \phi + \ell \right), B(u) = c u^{-2} \cos \phi, C(u) = c u^{-2} \left( -\sin \phi + \ell \right),
\end{equation}
where $\phi = 2\gamma \ln |u|$, and $c, \ell$ and $\gamma$ are constants. From Eqs.(\ref{ceq3-pw})-(\ref{ceq6-pw}) we find $c_3, c_4$ and $c_7$ vanish, and $c_6 = c_5 /\gamma$. Thus there is only one additional KV as
\begin{eqnarray}
& & {\bf X}_6 = \gamma ( z \p_y - y \p_z ) + u \p_u - v \p_v,
\end{eqnarray}
which yields the spacetime admits $\mathcal{N}_{9} \supset \mathcal{H}_7 \supset \mathcal{G}_6$.

(iii) {\it Isometry class 13} of \cite{sg}. In this case $H = (a y^2 + c z^2)/2 + b y z$, i.e. $A(u) =a, B(u)= b$ and $C(u) =c$, where $a, b,c$ are constants. For this case there exists only one additional KV as
\begin{eqnarray}
& & {\bf X}_6 =  \p_u.
\end{eqnarray}
Thus the spacetime admits $\mathcal{N}_{9} \supset \mathcal{H}_7 \supset \mathcal{G}_6$.

(iv) {\it Isometry class 14} of \cite{sg}. In this class,
\begin{equation}
A(u) = c \left( \sin \phi + \ell \right), \quad B(u) = c \cos \phi, \quad C(u) = c \left( -\sin \phi + \ell \right),
\end{equation}
where $\phi = 2 \gamma u$. Using Eqs.(\ref{ceq3-pw})-(\ref{ceq6-pw}), we get that $c_3, c_4, c_6$ vanish and $c_7 = c_5 / \gamma$ which give the additional KV
\begin{eqnarray}
& & {\bf X}_6 =  \gamma (z \p_y -y \p_z ) + \p_u.
\end{eqnarray}

\subsection{Conformally flat spacetimes}

Since $dim \mathcal{H} = 7$ for any conformally flat plane wave spacetime, it follows that the minimum dimension of NGS algebra for conformally flat plane wave spacetimes is \emph{nine}. We now give NGSs corresponding to the plane wave symmetry classes. Because of that $A(u) = C(u)$ and $B(u) = 0$ in conformally flat plane wave spacetimes which is just \emph{isometry class 15} of \cite{sg}, it follows from the constraint equations (\ref{ceq3-pw})-(\ref{ceq6-pw}) that $c_5$ is an arbitrary constant, $c_3=0$ and $c_4=0$. Therefore such spacetimes admit another KV given by
\begin{eqnarray}
& & {\bf X}_5 =  z \partial_y - y \partial_z .
\end{eqnarray}
This conformally flat pp-wave spacetimes, where the function $H$ has the form $H= \frac{1}{2} A(u) \delta_{AB} x^A x^B (A,B =2,3)$, may be physically considered as an Einstein-Klein-Gordon or Einstein-Maxwell solution \cite{mm}. Furthermore there are two specializations of the function $A(u)$.

(v) {\it Isometry class 16} of \cite{sg}. The function of $A(u)$ is a constant. Then the additional KV is
\begin{eqnarray}
& & {\bf X}_6 =  \partial_u.
\end{eqnarray}

(vi) {\it Isometry class 17} of \cite{sg}. In this case, the form of function $A(u)$ is $A(u) = a u^{-2}$. Thus we find the additional KV as
\begin{eqnarray}
& & {\bf X}_6 =  u \partial_u -v \p_v.
\end{eqnarray}
Thus we conclude that the maximum dimension of NGS algebra for conformally flat plane wave spacetimes is \emph{ten}.

\section{Discussions and Conclusions}
\label{CONC}

In this work we have examined NGSs of the geodesic Lagrangian for pp-wave spacetimes according to  isometry classes appearing in references \cite{kt} and \cite{sg}, and found the maximum dimension of the associated NGS Lie algebras which are mostly listed in Table \ref{T2} and Table \ref{T3}. In all of the isometry classes, the NGSs of the geodesic Lagrangian have all KVs of corresponding pp-wave spacetime and additionally some new symmetry generators are associated with the Lagrangian. The additional NGSs are specially important because they can in most instances give new first integrals of the geodesic equations. We found that a type N pp-wave spacetime is at least admitting three NGSs. Thus, the number of NGSs for which the plane wave spacetimes are not included can be \emph{four, five, six, seven} and \emph{eight} (see Tables 1, 2 and 3). It is further proved that the KV algebra is not only subalgebra of NGS algebra but the HKV (if it exists) gives also the subalgebra of NGSs  \cite{tsamparlis1,tsamparlis2}. This study includes a lot of examples for the fact that the KVs and HKV together with symmetry such as \emph{$s \partial_s$} are always the symmetries of the Lagrangian for the geodesic equations of spacetimes and but not SCKVs or proper CKVs. We note here that a HKV alone is not a NGS in this study.

We have found that all plane wave spacetimes admit at least \emph{eight} dimensional NGS algebra. The maximum dimension of the NGS algebra on $M$ is \emph{ten} which includes the 7 KVs, 1 HKV and 2 non-special NGSs if $M$ is conformally flat. If the spacetime is non-conformally flat plane wave, then the possible dimensions of NGS algebras for plane wave spacetimes can be \emph{nine} or \emph{ten}.

Because of the fact $D_s I =0$, the associated  conserved charges related with each of the NGS generator ${\bf Y}$'s will be the Noether first integral $I$'s which are explicitly obtained in Table 1, Table 2 and Table 5. It is well known that the pp-wave solutions are the simplest gravitational waves which are not directly detected so far, and so the search of the exact solutions of the geodesic equations for the pp-wave spacetimes would be an interesting task.

Considering each of the NGS generators and related first integrals, we have discussed the solution of geodesic equation in the pp-wave spacetimes through this paper. For the isometry classes in standard coordinates, we showed that how the Noether constants give the solution of geodesic equations. Every classes in these coordinates have the same three Noether symmetries and so the same three Noether constants which give the analytical solution of the $u$. In isometry class $1i$, we also found the exact solution of the $y$. So for this class we have the solution of geodesic equations both $u$ and $y$. The found symmetries with NGS technique in classes $3, \, 4, \, 8$ and $8(\epsilon \neq 0)$ are not enough to solve the other geodesic equations beside $u$. We found either an implicit solution of a geodesic equation or a relation between Noether constants, which are simply not enough to get exact solution of geodesic equations. But in the classes $9$ and $Biv$, with a random selection of Noether constants, we get the analytical solutions of the geodesic equations via NGS and so Noether constants. Since $\kappa = 2 I_2 = 0$, the solution of the geodesic equations in isometry class $9$ is a solution of a massless particle and because of $\kappa \neq 0$, the obtained solution in $Biv$ is for a massive particle.  We need to remind that these two analytical solutions of geodesic equations came with an arbitrary selection of Noether constants, but one can also find the another special solutions of these classes by using different combinations of the constants and that is a part of another study. Noether constants given in the rest of the classes except Table \ref{T7} needs to study separately.

\section*{Acknowledgements}

This work was supported by Akdeniz University, Scientific Research Projects Unit (BAP) under project number 2013.01.115.003.

\appendix

\section{Integration of Noether symmetry equations in standard coordinates}

The equations in (\ref{neq1234})-(\ref{neq15}) are an overdetermined PDE system. By solving these equations systematically, one can determine a general form for the components $\xi, \eta^1, \eta^2, \eta^3, \eta^4$ of Noether gauge symmetry generator ${\bf Y}$. It is easily solved from Eq. (\ref{neq-1})  that $\xi$ depends only on $s$, i.e. $\xi = \xi (s)$.
The first three equations given in (\ref{neq1234}) yield
\begin{eqnarray}
&& \eta^1 = A_1(s,u,y,z), \\
&& \eta^3 = v A_{1,y} + A_2(s,u,y,z), \\
&& \eta^4 = v A_{1,z} + A_3(s,u,y,z),
\end{eqnarray}
where $A_1, A_2, A_3$ are integration functions to be determined. Using the components $\eta^3$ and $\eta^4$ in the Eq. (\ref{neq567}), we find
\begin{eqnarray}
& & A_{1,yz}=0, \quad A_{1,yy} = 0, \quad A_{1,zz} = 0, \\
& & A_{2,y} - \frac{1}{2} \xi_{,s} = 0, \quad A_{2,z} - \frac{1}{2} \xi_{,s} = 0, \\
& & A_{2,z} + A_3{,y} = 0,
\end{eqnarray}
which have the following solution
\begin{eqnarray}
&& A_1 = y B_1(s,u) + z B_2(s,u) + B_3(s,u) \\
% && \eta^2 = v (\xi_s - y D_{1_{,u}} - z D_{2_{,u}} - D_{3_{,u}}) + A_4(s,u,y,z), \\
&& A_2 = \frac{y}{2} \xi_{,s} + K (s,u,z), \\
&& A_3 = \frac{z}{2} \xi_{,s} + L (s,u,y), \\
&& K_{,z} + L_{,y} = 0.\label{KL}
\end{eqnarray}
Then, the first equation of (\ref{neq89}) gives
\begin{equation}
\eta^2 = v \left( \xi_{,s} - A_{1,u} \right) + A_4 (s,u,y,z),
\end{equation}
where $A_4$ is an integration function. Substituting the obtained components $\eta^1, \eta^2, \eta^3$ to second and third equations of (\ref{neq89}), one can find that $B_{1,u} = B_{2,u} = 0$, that is, $B_1 =B_1(s)$ and $B_2 = B_2 (s)$, and
\begin{eqnarray}
& & A_{4,y} - K_{,u} + 2 B_1 H = 0, \\ & & A_{4,z} - L_{,u} + 2 B_2 H = 0.
\end{eqnarray}
When the potential $U(u,v,y,z)$ vanishes, the Eq. (\ref{neq15}) gives rise to that the gauge function $f$ depends only on the standard coordinates, i.e. $f=f(u,v,y,z)$. Now, it remains the Eqs. (\ref{neq10-13}) and (\ref{neq14}) to be solved. Thus, using the obtained components in Eq. (\ref{neq10-13}) and then considering the Eq.(\ref{KL}), we find
\begin{eqnarray}
& & B_1 = a_1, \quad B_2 = a_2, \quad B_3 = a_3 s  + k_1 (u),  \\ & & K = s k_2 (u) + z k_3 (u) + k_4 (u), \quad  L = s k_5 (u) - y k_3 (u) + k_6 (u), \\ & & \xi = c_1 s + c_2, \\& &f = -a_3 v + y k_2 (u) + z k_5 (u) + k_7 (u)
\end{eqnarray}
with the constraint equations for $A_4$
\begin{eqnarray}
& & A_{4,y} - s k_{2,u} - z k_{3,u} - k_{4,u} + 2 a_1 H = 0, \\ & & A_{4,z} - s k_{5,u} + y k_{3,u} - k_{6,u} + 2 a_2 H = 0, \\& & A_{4,s} + y k_{2,u} + z k_{5,u} + k_{7,u} + 2 a_3 H = 0,
\end{eqnarray}
where $k_1, ..., k_7$ are integration functions, and $a_1, a_2, a_3, c_1,c_2$ are integration constants. Finally, we consider the Eq. (\ref{neq14}) with the above results and find that $a_1, a_2$ and $a_3$ vanish, and
\begin{eqnarray}
& & k_1 = c_6 u + c_7, \quad k_2 = c_3, \quad k_3 = c_5, \quad k_5 = c_4, \\ & & A_4 = -s f_{1,u} + y f_{2,u} + z f_{3,u} + f_4 (u),
\end{eqnarray}
where $f_4(u)$ is an integration function, and $c_3,..., c_7$ are constants of integration. In order to get the same form of solutions given by (\ref{xieta1})-(\ref{f}) with constraint conditions (\ref{ceq1}) and (\ref{ceq2}), we have changed the functions $k_7(u), k_4(u)$ and $k_6 (u)$ to $f_1 (u), f_2 (u)$ and $f_3(u)$, respectively.

%\section*{}

\end{document}